\documentclass[preprint, 12pt]{elsarticle}

\usepackage{fullpage}
\usepackage{array}
\usepackage{algorithm}
\usepackage{algpseudocode}
\usepackage{pifont}
\usepackage{bbding}
\usepackage{graphicx}
\usepackage{subfig}
\usepackage{enumerate}
\usepackage[shortlabels]{enumitem}
\usepackage{amsmath,amssymb,mathtools}
\usepackage{amsfonts,bm}
\usepackage{todonotes}
\usepackage[commentmarkup=footnote]{changes}
\usepackage{stmaryrd}
\usepackage{siunitx}
\usepackage{caption}
\usepackage{array}
\usepackage{booktabs}
\usepackage{makecell}
\usepackage{nomencl}
\usepackage{lineno}
\usepackage{url}
\usepackage{natbib}
\usepackage[dvipsnames]{xcolor}
\usepackage[strings]{underscore}
\usepackage[commentmarkup=footnote]{changes}
\definecolor{darkgreen}{rgb}{0.0, 0.4, 0.0}
\definechangesauthor[name={Reviewer 1}, color = darkgreen]{R1}
\definechangesauthor[name={Reviewer 2}, color = red]{R2}
\definechangesauthor[name={Author}, color = blue]{Author}

\journal{}

\begin{document}

\begin{frontmatter}

\title{Neural ensemble Kalman filter: Data assimilation for\\compressible flows with shocks}

\author[sd,cal]{Xu-Hui Zhou\corref{cor1}}%
\ead{xuz067@ucsd.edu}
\author[sd]{Lorenzo Beronilla}%
\author[cal]{Michael K. Sleeman}%
\author[jh]{Hangchuan Hu}%
\author[sd]{\\Matthias Morzfeld}%
\author[cal]{Andrew M. Stuart}%
\author[jh]{Tamer A. Zaki}%

\cortext[cor1]{Corresponding author.}

\affiliation[sd]{organization={Scripps Institution of Oceanography, University of California San Diego},
            city={La Jolla},
            postcode={92093}, 
            state={CA},
            country={USA}}
\affiliation[cal]{organization={Computing + Mathematical Sciences, California Institute of Technology},
            city={Pasadena},
            postcode={91125}, 
            state={CA},
            country={USA}}
\affiliation[jh]{organization={Department of Mechanical Engineering, Johns Hopkins University},
            city={Baltimore},
            postcode={21218}, 
            state={MD},
            country={USA}}

\begin{abstract}
Data assimilation (DA) for compressible flows with shocks is challenging because many classical DA methods generate spurious oscillations and nonphysical features near uncertain shocks.  We focus here on the ensemble Kalman filter (EnKF). We show that the poor performance of the EnKF may be attributed to the bimodal forecast distribution that can arise in the vicinity of an uncertain shock location; this violates the assumptions underpinning the EnKF, which assume a forecast which is close to Gaussian. To address this issue we introduce the new \emph{neural EnKF}.
The basic idea is to systematically embed neural function approximations within ensemble DA by mapping the forecast ensemble of shocked flows to the parameter space (weights and biases) of a deep neural network (NN) and to subsequently perform DA in that space.
The nonlinear mapping encodes sharp and smooth flow features in an ensemble of NN parameters. 
Neural EnKF updates are therefore well-behaved only if the NN parameters vary smoothly within the neural representation of the forecast ensemble. We show that such a smooth variation of network parameters can be enforced via physics-informed transfer learning, and demonstrate that in so-doing the neural EnKF avoids the spurious oscillations and nonphysical features that plague the EnKF.
The applicability of the neural EnKF is demonstrated through a series of systematic numerical experiments with the inviscid Burgers' equation, the Sod shock tube, and a two-dimensional blast wave. 
\end{abstract}

\begin{keyword}
data assimilation, ensemble Kalman filter, shocks, discontinuities, deep neural networks
\end{keyword}

\end{frontmatter}

\section{Introduction}
\label{sec:introduction}
Compressible flows featuring shocks and other discontinuities arise in a wide range of natural and engineered systems. In rotating detonation engines, a continuously propagating detonation wave circles the annular combustor, producing strong shocks essential for pressure-gain combustion~\cite{raman2023nonidealities}; in high-speed flight, shock waves form at the leading edges of flight vehicles and at sharp compression corners, strongly influencing the resulting pressure and thermal loads~\cite{anderson1989hypersonic,schuabb2025hypersonic}; and in shockwave and laser lithotripsy, a focused acoustic or laser-induced pulse initiates cavitation bubble collapse near kidney stones, producing localized shock emissions that contribute to stone fragmentation~\cite{cao2021shock,zhao2024vapour}. These examples, among many other applications~\cite{cao2019shock, ma2022computational, narkhede2025fluid}, highlight the significant role of shocks in modern science and engineering and underscore the importance of accurately simulating compressible flows with shocks. When the flow conditions (e.g., initial and boundary conditions) are well characterized, numerical solvers may produce high-fidelity predictions. In practice, however, significant uncertainties often arise from incomplete or imperfectly specified conditions, motivating the use of data-driven methods in conjunction with numerical simulations to reduce uncertainty and improve predictive capability.

Data assimilation (DA) provides a systematic Bayesian framework for reducing these uncertainties~\cite{evensen2009data,asch2016data,evensen2022data}. In many DA formulations, the system state is sequentially updated as new observations become available: with each observation, a prior estimate predicted by a numerical model is merged with observation data to form a posterior state that more faithfully represents both the governing physics and the observations.
DA was initially developed within the field of numerical weather prediction~\cite{kalnay2003atmospheric}, where atmospheric observations are used to keep atmospheric models on track, and it has since become widely adopted throughout the geosciences~\cite{carrassi2018data}.
Building on this success, DA has seen growing application in computational fluid dynamics and related engineering problems to improve modeling fidelity and enhance predictive accuracy under uncertainty (e.g.,~\cite{zhang2022ensemble,zhou2023inference,zafar2026data,liu2025towards,wang2021state,zaki2025turbulence}).
However, most existing studies have focused on incompressible flows with smooth features. 
Effective DA for compressible \emph{shocked} flows is an emerging sub-field~\cite{buchta2021observation,buchta2022assimilation,morra2024ml,wang2025domain}.

Existing work on DA for compressible shocked flows can be broadly categorized into variational (smoothing) and sequential (filtering) methods. 
Within the variational paradigm, \citet{west2025variational} applied a 4D (3D space and time) variational framework (4D-Var) to the Sod shock tube problem, estimating initial conditions that yield a solution trajectory best matching the observations over a given time interval. 
In contrast to identifying an optimal initial condition, sequential filtering approaches update the flow state dynamically as observations become available.
\citet{houba2024sequential} presented a systematic investigation of sequential filtering approaches for compressible flows with shocks, evaluating extended Kalman filter (EKF), ensemble Kalman filter (EnKF), and particle filter (PF) on the Sod shock tube problem. Their results showed that, due to uncertainty in shock locations, both the EKF and EnKF perform poorly in shock-dominated regimes, with the EnKF exhibiting pronounced oscillations and violations of thermodynamic realizability. In contrast, PF provided the most robust performance, provided that a sufficiently rich and appropriately structured initial ensemble is available.
In subsequent work, \citet{edoh2025sequential} introduced a positivity-preserving logarithmic variable transformation to enforce the thermodynamic realizability in the EnKF update; however, spurious oscillations near shocks were not eliminated.
\citet{edoh2026data} further demonstrated that these oscillatory artifacts are significantly reduced when the initial discontinuity locations are assumed to be accurately known without uncertainty.

To further improve the EnKF for shocked flows, \citet{hansen2024normal} applied a normal-score EnKF, originally proposed by~\citet{zhou2011approach}, to the Sod shock tube problem. 
The method applies a rank-based transformation to map the distribution at each spatial location toward a Gaussian distribution before the EnKF update. While this improved the performance compared to the EnKF, spurious oscillations near shocks remained because Gaussian marginal distributions do not necessarily imply a Gaussian joint distribution.
They further investigated DA in the context of detonation wave dynamics and proposed a two-step local ensemble transform Kalman filter (LETKF) framework that aligns ensemble members based on the detonation front prior to the analysis step~\cite{hansen2025data}.
In parallel, \citet{li2024structurally,li2025structurally} proposed a structurally informed ETKF method that incorporates weighting matrices derived from state gradients into the prior covariance, thereby explicitly encoding local discontinuity information in the analysis step. This method was demonstrated for several nonlinear hyperbolic problems, including the dam-break problem, the shallow water equations, and the 2D Burgers’ equation, under relatively dense observation settings. However, its performance for compressible shocked flows under sparse observations has not yet been investigated.
Beyond ensemble Kalman filtering approaches, \citet{subrahmanya2025feature} introduced a feature-preserving DA method based on the ensemble transform particle filter (ETPF) to address the fundamental issue of feature smearing during DA. The method formulates the analysis update as an optimal transport problem to determine the ensemble transform weights. These weights are subsequently decomposed into a sequence of pairwise convex combinations of particles, while dynamic time warping is applied to align particles pairwise in order to preserve sharp flow features in the posterior states. The method showed promising results on test cases including shock tube problems and a 2D blast wave case.
Nevertheless, it relies on the presence of consistent features across particles, 
and can suffer from particle collapse unless inflated observation errors are used.
In a different line of PF-based approaches, \citet{srivastava2023feature} proposed a feature-informed DA method that directly assimilates solution features, such as shock locations, to infer a small set of governing parameters instead of performing full-state estimation. However, a small set of parameters is often insufficient to capture the dynamics of practical shock-laden flows.

In this work, we analyze the challenges associated with applying the stochastic EnKF to compressible flows with shocks and propose a simple yet effective approach. We show that the degradation of the EnKF near shocks can be traced to a fundamental mechanism: uncertainty in the shock locations induces strongly bimodal forecast distributions in physical space, violating the assumptions underlying the EnKF and leading to spurious oscillations and nonphysical updates. This mechanism is illustrated using a hyperbolic-tangent surrogate shock as a diagnostic example. Motivated by this insight, we propose performing EnKF updates in an alternative representation and develop the \emph{neural EnKF}, which carries out the analysis step in the neural network parameter space (hereafter referred to as the \emph{neural space}). In this approach, each ensemble member is represented by a deep neural network, with a shared architecture across all members, whose weights (and biases) implicitly encode the underlying flow features. The networks are trained progressively along a nearest-neighbor chain constructed from physical-space similarity to maintain well-behaved ensemble statistics in the neural space. Observations in physical space, such as pressure measurements, are assimilated by updating the network weights, yielding an analysis ensemble of networks whose reconstructed flow fields better reflect the data while preserving sharp flow features.
We assess the DA performance of the neural EnKF through a series of systematic numerical experiments, including the inviscid Burgers’ equation, the Sod shock tube problem, and a 2D blast wave case. These test cases demonstrate the neural EnKF’s ability to perform robust DA in shock-laden flows.

The remainder of this paper is organized as follows.
Section~\ref{sec:motivation} examines the behavior of the EnKF in the presence of uncertain shocks, using a hyperbolic-tangent example to illustrate how sharp transitions challenge the EnKF updates and to clarify the underlying failure mechanism.
Section~\ref{sec:method} presents the neural EnKF methodology, including the neural-network representation of ensemble members, the EnKF update in neural space, and a nearest-neighbor chain training strategy designed to maintain well-behaved ensemble statistics in the neural space.
Section~\ref{sec:result} evaluates the proposed method through a series of systematic numerical experiments in shock-laden flows.
Finally, Section~\ref{sec:conclusion} summarizes the main findings and discusses directions for future research.

\section{Motivation}
\label{sec:motivation}
We begin by briefly reviewing the stochastic EnKF~\cite{burgers1998analysis}, which is a natural choice for the problems considered here, because it is derivative free and because it scales to high-dimensional systems; these two features have led to its widespread adoption in many practical applications including weather prediction, climate science, physical oceanography and reservoir engineering. Despite these advantages, the EnKF exhibits notable difficulties when applied to compressible flows with shocks. To study this behavior, we use a hyperbolic-tangent surrogate shock to examine how sharp transitions challenge the EnKF update and lead to spurious oscillations.

\subsection{Ensemble Kalman filter (EnKF)}
The EnKF updates a forecast ensemble, predicted by a numerical model, using observations. At each DA cycle, forward simulations generate a forecast ensemble 
$\{\mathbf{z}_i^{\mathrm{f}}\}_{i=1}^{n_e}$, where $\mathbf{z}_i^{\mathrm{f}} \in\mathbb{R}^{n_z}$ denotes the forecast state of the $i$th ensemble member and $n_e$ is the ensemble size.
The ensemble is characterized by its mean and normalized ensemble perturbations:
\begin{equation}
    \bar{\mathbf{z}}^{\mathrm{f}}
    = \frac{1}{n_e} \sum_{i=1}^{n_e} \mathbf{z}_i^{\mathrm{f}},
    \qquad
    \mathbf{Z} = \frac{1}{\sqrt{n_e - 1}}
    \left[
        \mathbf{z}_1^{\mathrm{f}} - \bar{\mathbf{z}}^{\mathrm{f}},
        \ldots,
        \mathbf{z}_{n_e}^{\mathrm{f}} - \bar{\mathbf{z}}^{\mathrm{f}}
    \right]
    \in\mathbb{R}^{n_z \times n_e}.
\end{equation}
Predicted observations are defined as 
$\mathbf{y}_i = \mathcal{H}(\mathbf{z}_i^{\mathrm{f}})$, 
where $\mathcal{H}$ is the observation operator that maps the state space to the observation space.
The ensemble of predicted observations is characterized by its mean and normalized ensemble perturbations:
\begin{equation}
    \overline{\mathbf{y}}
    = \frac{1}{n_e}\sum_{i=1}^{n_e}\mathbf{y}_i, 
    \qquad
    \mathbf{Y}
    =
    \frac{1}{\sqrt{n_e - 1}}
    \left[
        \mathbf{y}_1 - \overline{\mathbf{y}},
        \ldots,
        \mathbf{y}_{n_e} - \overline{\mathbf{y}}
    \right] \in\mathbb{R}^{n_y \times n_e}.
    \label{eq:y-mean-anom}
\end{equation}
The EnKF combines the forecast ensemble, the predicted observations, and the observations $\mathbf{d}$ to obtain the analysis ensemble members:
\begin{equation}
    \mathbf{z}_i^{\mathrm{a}} = \mathbf{z}_i^{\mathrm{f}} +
    \mathbf{Z}\mathbf{Y}^{\top}
    \left( \mathbf{Y}\mathbf{Y}^{\top} + \mathbf{R} \right)^{-1}
    \left( 
        \mathbf{d} + \bm{\eta}_i - \mathbf{y}_i
    \right),
    \qquad
    \bm{\eta}_i \sim \mathcal{N}(\mathbf{0},\mathbf{R}),
    \label{eq:enkf-update}
\end{equation}
where $\mathbf{R} \in \mathbb{R}^{n_y \times n_y}$ is the observation error covariance (a symmetric positive semidefinite matrix), and $\bm{\eta}_i$ represents the independent observation perturbations.
We focus here on the stochastic EnKF as a representative EnKF formulation, although the ideas developed in this work can be naturally extended to other variants.

The EnKF update shows that each analysis member is obtained as a linear combination of the forecast ensemble. As a result, the analysis ensemble remains within the subspace spanned by the forecast ensemble, a feature commonly referred to as the \emph{subspace property} of the EnKF~\cite{evensen2022data,iglesias2013ensemble}.
This subspace property can become restrictive when reconstructing functions with sharp gradients using (approximately) discontinuous basis functions. We illustrate this effect through a simple example in the following subsection.

\subsection{Ensemble Kalman filtering in the presence of a ``shock''}
We investigate the behavior of the EnKF in the presence of a sharp, shock-like transition using a diagnostic example based on a hyperbolic-tangent profile,
\begin{equation}
    h(x)=\frac{h_L + h_R}{2}
    -\frac{h_L - h_R}{2}
    \tanh\!\left(\kappa\left(x-x_s\right)\right),
    \label{eq:tanh-func}
\end{equation}
where $h_L$ and $h_R$ denote the left and right plateau values, $\kappa > 0$ controls the steepness of the transition, and $x_s$ specifies the nominal shock location. This smooth yet rapidly varying profile provides a controlled approximation of a 1D shock.

\begin{figure}[!htb]
\centering
\includegraphics[width=0.99\textwidth]{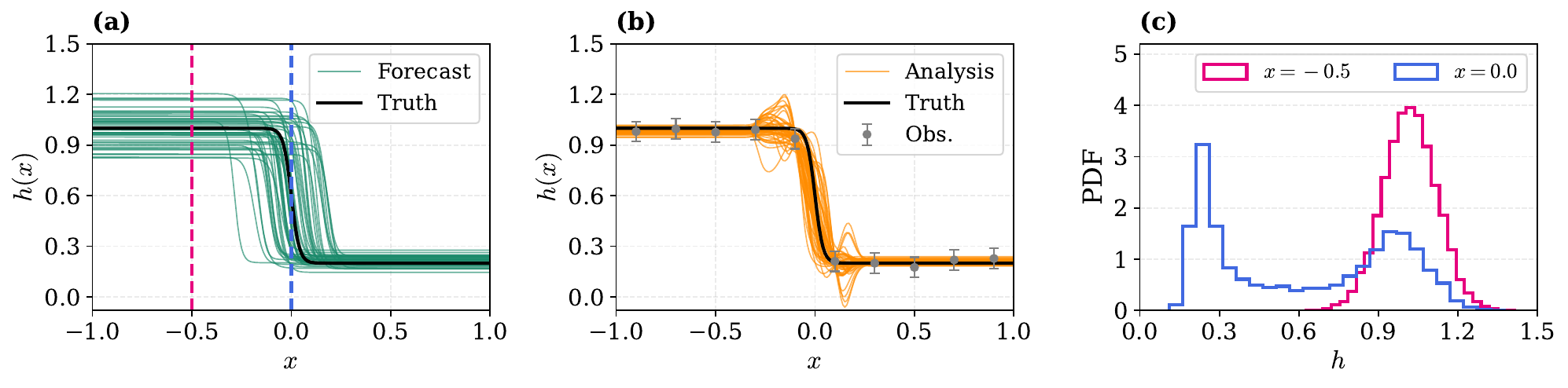}
  \caption{Behavior of the EnKF for a shock-like transition, illustrated using a hyperbolic-tangent example: 
  (a) forecast ensemble with 50 members; 
  (b) analysis ensemble obtained using the EnKF, showing spurious oscillations near the shock; and 
  (c) marginal probability density functions (PDFs) of the forecast ensemble at two representative locations indicated in (a), showing an approximately Gaussian distribution away from the shock ($x=-0.5$) and a bimodal distribution near the shock location ($x=0$).
  For clarity, the PDFs in (c) are estimated using 10,000 forecast ensemble
  members.
  Here, the error bars represent $\pm 2$ standard deviations of the observational noise; the same convention is adopted in all subsequent figures containing error bars.}
  \label{fig:tanh-EnKF}
\end{figure}

The true state $\mathbf{z}^*$ is defined by the parameter set $(h_L, h_R, \kappa, x_s) = (1.0, 0.2, 25, 0)$ and evaluated on a uniform grid over $x \in [-1,1]$ with spacing $\Delta x = 0.01$. To represent uncertainty, a 50-member forecast ensemble is generated by independently sampling the parameters $(h_L, h_R, \kappa, x_s)$ 
from (truncated) Gaussian distributions:
\begin{equation}
h_L \sim \mathcal{N}(1.02,\,0.1^2),\quad
h_R \sim \mathcal{N}(0.22,\,0.03^2),\quad
\kappa \sim \mathcal{N}_{(0,\infty)}(25,\,4.0^2),\quad
x_s \sim \mathcal{N}_{(-1,1)}(0,\,0.1^2).
\label{eq:tanh-sampling}
\end{equation}
Here, $\mathcal{N}_{(a,b)}(\mu,\sigma^2)$ denotes a Gaussian distribution with mean $\mu$ and variance $\sigma^2$ truncated to the interval $(a,b)$, with the truncation enforced through rejection sampling.
Each sampled parameter set defines a forecast ensemble member, forming the forecast ensemble shown in Fig.~\ref{fig:tanh-EnKF}(a).

Synthetic observations are constructed by sampling the true state $\mathbf{z}^*$ at 10 uniformly spaced locations with spacing $\Delta x_\mathrm{obs} = 0.2$, located at
$x = \pm\, 0.1, \pm\, 0.3, \ldots, \pm\, 0.9$, and adding independent Gaussian noise,
\begin{equation}
    \mathbf{d} = \mathsf{H}\mathbf{z}^* + \bm{\eta},
    \qquad
    \bm{\eta} \sim \mathcal{N}(\mathbf{0}, 0.03^{2}\mathbf{I}),
    \label{eq:tanh-observation}
\end{equation}
where $\mathsf{H}$ denotes a linear observation operator that extracts the state values at the observation locations.
Consistent with this observation model, the predicted observations for each forecast ensemble member $\mathbf{z}_i^{\mathrm{f}}$ are given by
\begin{equation}
    \mathbf{y}_i
    = \mathsf{H} \mathbf{z}_i^{\mathrm{f}}.
    \label{eq:tanh-predicted}
\end{equation}
Given the forecast ensemble $\{\mathbf{z}_i^{\mathrm{f}}\}_{i=1}^{50}$, the predicted observations $\{\mathbf{y}_i\}_{i=1}^{50}$, and the synthetic data $\mathbf{d}$, the EnKF update is performed according to Eq.~\eqref{eq:enkf-update}.

The resulting analysis ensemble exhibits pronounced oscillations in the vicinity of the sharp transition, as illustrated in Fig.~\ref{fig:tanh-EnKF}(b). This behavior reflects a fundamental limitation of the EnKF: it performs well when the forecast state distribution is Gaussian or nearly Gaussian, but becomes ineffective in strongly non-Gaussian scenarios~\cite{morzfeld2019gaussian}. In this example, uncertainty in the shock location induces pronounced non-Gaussianity. Variations in the shock position cause ensemble members to place the steep transition at different locations, such that some members have already transitioned to the right plateau while others remain on the left at a given spatial point.
This effect is further illustrated in Fig.~\ref{fig:tanh-EnKF}(c), which shows the marginal distributions of the forecast ensemble at two representative locations: one in a smooth region far from the shock ($x=-0.5$) and one near the shock ($x=0$). At $x=-0.5$, the distribution is approximately Gaussian. In contrast, at $x=0$, the distribution is clearly bimodal, with one mode corresponding to ensemble members whose shock lies to the left of the point and another corresponding to those whose shock lies to the right. This pronounced bimodality explains why the EnKF produces spurious oscillations when applied near discontinuities.
Such spurious oscillations are undesirable in compressible shock-laden flows, as they severely distort the physical structure of the solution. 

\section{Methodology}
\label{sec:method}
We introduce a new variant of an EnKF, the neural EnKF, to address problems caused by bimodal forecast distributions common in shock-laden flows (see above).
The basic idea is to embed neural function approximations in the EnKF by mapping the forecast ensemble to the parameter space (weights and biases) of a deep neural network (NN) and to subsequently perform DA in that space.
The nonlinear mapping encodes sharp and smooth flow features in a transformed ensemble of NN parameters. 
A key challenge here is associated with the NN parameterization of the forecast ensemble. This challenge motivates a transfer learning strategy that enforces smooth variation in the neural space through a nearest-neighbor chain.

\subsection{Neural EnKF}

\begin{figure}[!htb]
\centering
\includegraphics[width=0.99\textwidth]{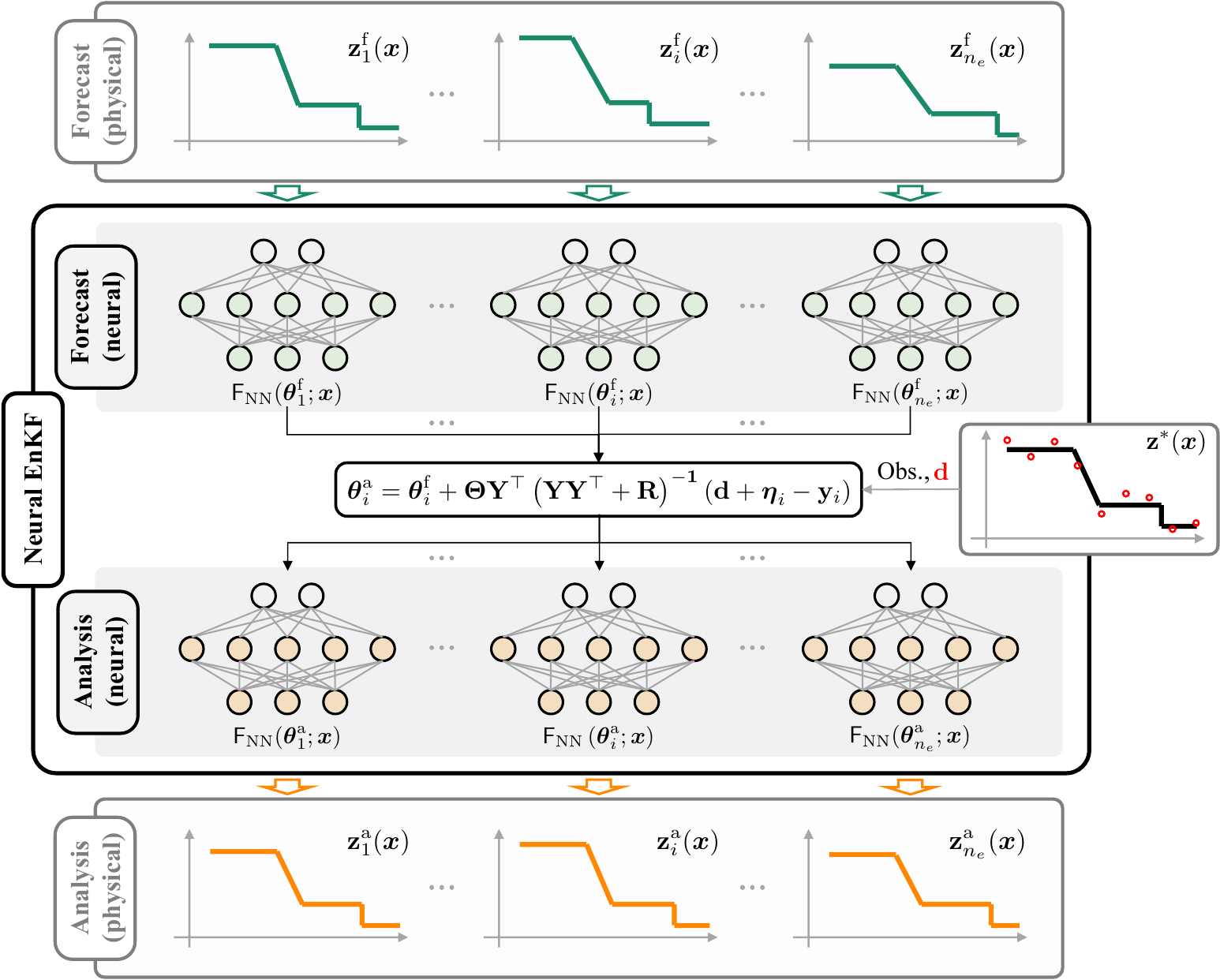}
\caption{
Schematic of the neural EnKF framework. At each DA cycle, forward simulations generate a forecast ensemble $\{\mathbf{z}_i^{\mathrm{f}}(\bm{x})\}_{i=1}^{n_e}$ in physical space. Each forecast ensemble member is then parameterized by a deep neural network $\mathsf{F}_{\mathrm{NN}}(\bm{\theta}_i^{\mathrm{f}}; \bm{x})$, which maps the spatial coordinates to the corresponding physical state (e.g., density, velocity, and pressure in compressible flows). The network parameters $\bm{\theta}_i^{\mathrm{f}}$ are then updated to $\bm{\theta}_i^{\mathrm{a}}$ through assimilation of the physical-space observations $\mathbf{d}$ (e.g., pressure measurements). Finally, the updated networks reconstruct the analysis ensemble $\{\mathbf{z}_i^{\mathrm{a}}(\bm{x})\}_{i=1}^{n_e}$ back in physical space.}
\label{fig:neural-EnKF}
\end{figure}

The neural EnKF consists of three main steps, as illustrated in Fig.~\protect\ref{fig:neural-EnKF}: (i) neural-network parameterization of the forecast ensemble members, (ii) the EnKF update in the neural space, and (iii) reconstruction of the analysis ensemble in physical space.

Specifically, at each DA cycle, forward simulations generate a forecast ensemble $\{\mathbf{z}_i^{\mathrm{f}}\}_{i=1}^{n_e}$ on the spatial grid $\bm{x}$.  
Each forecast state $\mathbf{z}_i^{\mathrm{f}}$ is then mapped into a neural representation by training a deep neural network $\mathsf{F}_{\mathrm{NN}}(\bm{\theta}_i^{\mathrm{f}})$, where $\bm{\theta}_i^{\mathrm{f}}$ denotes the network parameters.
In this work, each ensemble member is represented by a single fully connected feedforward network, which predicts all flow variables simultaneously from the spatial coordinates.
In the final layer, softplus activations are applied to the thermodynamic outputs (e.g., density and pressure) to ensure positivity and improve realizability of the reconstructed physical states.
Each network is trained in a standard supervised learning manner to represent the corresponding forecast state by minimizing a mean-squared reconstruction loss.
This procedure yields a forecast ensemble of neural networks with identical architectures but distinct network parameters $\bm{\theta}_i^{\mathrm{f}}$.

With the neural representation established, the EnKF update is performed in the neural space, where the forecast ensemble is characterized by the mean and normalized perturbations of the network parameters:
\begin{equation}
\overline{\bm{\theta}}^{\mathrm{f}} = \frac{1}{n_e}\sum_{i=1}^{n_e}\bm{\theta}_i^\mathrm{f},
\qquad
\mathbf{\Theta} = \frac{1}{\sqrt{n_e-1}}
\left[\,\bm{\theta}_1^\mathrm{f}-\overline{\bm{\theta}}^\mathrm{f},\;\ldots,\;
\bm{\theta}_{n_e}^\mathrm{f}-\overline{\bm{\theta}}^\mathrm{f}\,\right]
\in\mathbb{R}^{n_\theta \times n_e}.
\label{eq:theta-mean-anom}
\end{equation}
The corresponding predicted observations are computed as $\mathbf{y}_i = \mathcal{H}(\mathbf{z}_i^{\mathrm{f}})$. Since the neural networks in this work are trained to approximate the physical states with fitting errors that are negligible relative to the prescribed observation error, applying the observation operator to the physical states or to the neural network reconstructions yields essentially identical results.
The resulting ensemble of predicted observations is characterized by its mean and normalized ensemble perturbations as in Eq.~\eqref{eq:y-mean-anom}.
The EnKF update is then applied to the network parameters,
\begin{equation}
\bm{\theta}_i^{\mathrm a}
=
\bm{\theta}_i^{\mathrm f}
+
\mathbf{\Theta}\mathbf{Y}^\top
\left(\mathbf{Y}\mathbf{Y}^\top + \mathbf{R}\right)^{-1}
\left(\mathbf{d} + \bm{\eta}_i - \mathbf{y}_i\right),
\qquad
\bm{\eta}_i \sim \mathcal{N}(\mathbf{0},\mathbf{R}),
\label{eq:neural-enkf-update}
\end{equation}
yielding an analysis ensemble of network parameters $\{\bm{\theta}_i^{\mathrm a}\}_{i=1}^{n_e}$.
The corresponding analysis states in physical space are then obtained through reconstruction,
\begin{equation}
\mathbf{z}_i^{\mathrm a}(\bm{x})
= \mathsf{F}_\mathrm{NN}(\bm{\theta}_i^{\mathrm a}; \bm{x}).
\label{eq:reconstruction}
\end{equation}

One can view the mapping from forecasted physical states to neural-network parameters as a \emph{nonlinear} coordinate transformation applied prior to the EnKF analysis step.
The nonlinearity of this transformation is important, because the affine invariance of the EnKF implies that filter performance is unaffected by linear or affine transformations.

We note that the neural EnKF differs from several existing latent-space DA approaches in important ways. Approaches such as those in~\cite{peyron2021latent,mousavi2025sequential} primarily use NNs to construct reduced-order latent representations and surrogate dynamics for efficient forecast evolution within DA. In contrast, this work is directly motivated by the analysis step itself. Rather than replacing the forward solver or evolving latent dynamics, NNs are used to learn representations of compressible flows with shocks in which ensemble-based Bayesian updates become more reliable. In this sense, our work shares similarities with recent efforts aimed at learning latent representations more compatible with Kalman-based filtering~\cite{tong2026latent}.
Our work also differs from recent studies aiming to improve the analysis step, where NNs are utilized either to enhance specific terms in the EnKF update formula~\cite{bach2025learning} or to directly learn the analysis mapping~\cite{zhou2024bi}, while still performing the analysis step in the original state space.

\subsection{Navigating non-convex loss landscapes with transfer learning}

While the neural EnKF provides a neural-space formulation for updating ensemble members, constructing the forecast ensemble in this space requires training a separate neural network for each member. This introduces an additional challenge: neural networks used to represent flow fields are typically overparameterized, resulting in highly non-convex optimization landscapes with many equivalent or near-equivalent minima. As a result, independently trained networks can converge to disparate locations in neural space despite representing nearly identical physical states.

\begin{figure}[tb]
\centering
\includegraphics[width=0.95\textwidth]{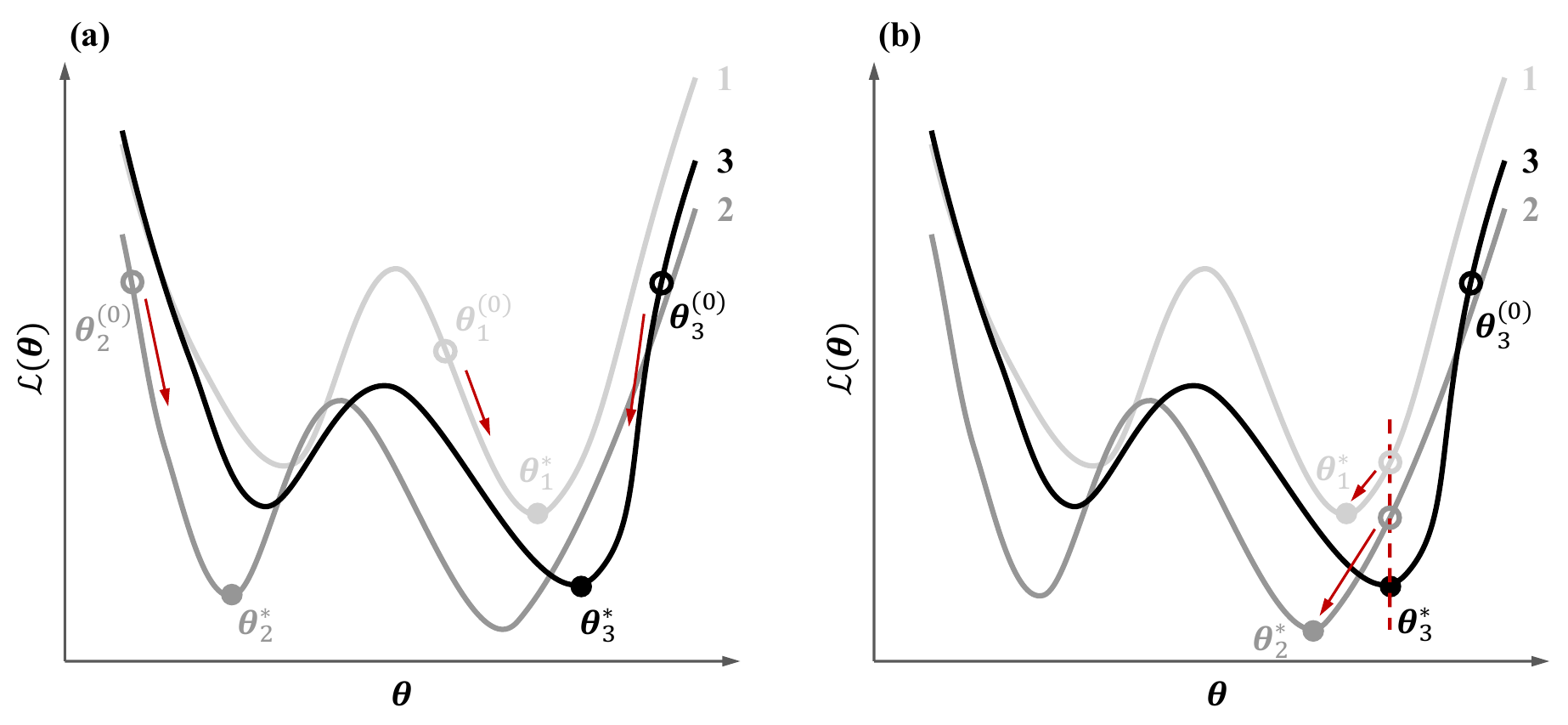}
    \caption{Schematic illustration of neural-network training across ensemble members on a non-convex loss landscape.
    (a) Independent training: each ensemble member is initialized from a different random starting point $\bm{\theta}_i^{(0)}$ and optimized independently, leading to convergence toward distinct local minima $\bm{\theta}_i^\ast$ and a poorly aligned ensemble in neural space (e.g., ensemble members 1 and 3 converge to nearby minima, whereas member 2 converges to a separated minimum).
    (b) Transfer learning: ensemble members are initialized from the trained parameters of a nearby ensemble member (e.g., $\bm{\theta}_3^\ast$), guiding successive optimization processes toward neighboring regions of the loss landscape and promoting alignment of ensemble parameters in neural space.}
    \label{fig:loss-landscape}
\end{figure}

This behavior is illustrated schematically in Fig.~\ref{fig:loss-landscape}(a) using three ensemble members as an example, where each curve represents the loss landscape associated with one ensemble member. Under independent training, each network is initialized from a distinct random parameter vector $\bm{\theta}_i^{(0)}$ and optimized separately, following its own trajectory toward a local minimum $\bm{\theta}_i^*$ in the neural space. Owing to the highly non-convex loss landscape, these minima may be widely separated, even when the corresponding physical states are nearly identical. As a result, the ensemble of trained networks exhibits poorly aligned parameters, which distorts the ensemble covariance structure in neural space and undermines the effectiveness of subsequent neural EnKF updates.

To address this issue, we introduce a transfer learning strategy, as illustrated in Fig.~\ref{fig:loss-landscape}(b). In this approach, the training of each ensemble member is initialized using the trained parameters of a nearby ensemble member (e.g., $\bm{\theta}_3^\ast$), thereby encouraging successive optimization processes to converge toward neighboring local minima. While this behavior cannot be guaranteed in a non-convex setting, the approach promotes parameter alignment in neural space by reducing divergence arising from random initialization.
As a result, the ensemble of trained networks exhibits a more coherent organization in neural space, such that variations among ensemble members more faithfully reflect differences in the underlying physical states rather than artifacts of independent optimization. This improved alignment preserves a meaningful ensemble covariance structure, which is essential for stable and effective neural EnKF updates.

\subsection{Nearest-neighbor chain training}
We formalize the concept of transfer learning through a nearest-neighbor chain training strategy, in which the ensemble of neural networks is trained sequentially along a chain constructed based on physical-space similarities among ensemble members.

\begin{figure}[!htb]
\centering
\includegraphics[width=0.99\textwidth]{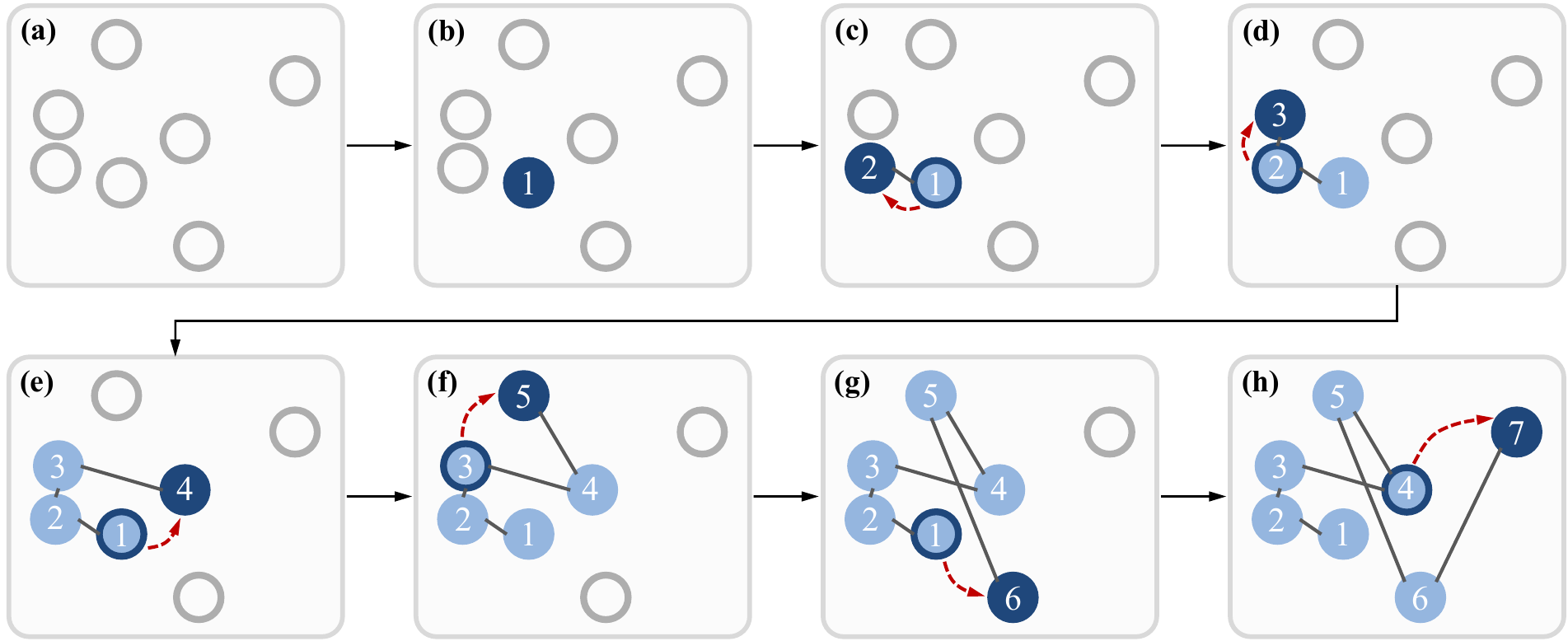}
    \caption{Schematic illustration of the nearest-neighbor chain construction:
    (a) ensemble members in the chosen metric space;
    (b) initialization of the chain by selecting the most ``central'' ensemble member, defined as the member with the smallest average distance to all others; and
    (c)--(h) iterative construction of the chain, where at each step the ensemble member closest to the currently \textbf{selected set} is appended.
    Unselected members are shown as grey circles, previously selected members in light blue, and the member added at the current step is highlighted in dark blue.
    Solid/black lines indicate the resulting nearest-neighbor chain, with numbers denoting the order of insertion.
    Dashed/red arrows indicate the parent–child relationship used for training initialization, pointing from the nearest previously selected member (shown in light blue with a dark outline) to the newly added member.}
    \label{fig:chain-construct}
\end{figure}

The construction of the nearest-neighbor chain is illustrated schematically in Fig.~\ref{fig:chain-construct}. Starting from the initial ensemble configuration (panel a), the construction begins by selecting the most ``central'' ensemble member (panel b), defined as the one with the smallest average distance to all other members under the chosen metric. The chain is then built iteratively by appending, at each step, the ensemble member that is closest to the currently \textbf{selected set} (panels c-h). Repeating this procedure yields an ordered chain that links all ensemble members and defines the sequence in which they are trained.
It is important to note that the training initialization for a newly added ensemble member is not inherited from its immediate predecessor in the chain, but from its nearest neighbor within the set of previously selected members.
For example, in panel (e), the newly added ensemble member 4 is initialized from member 1, its nearest neighbor within the selected set ${1,2,3}$, rather than from member 3, which immediately precedes it in the chain.
By enforcing this local nearest-neighbor dependency, the construction ensures that each network is initialized from the most relevant predecessor in physical space.

The nearest-neighbor chain construction described above is formalized in Algorithm~\ref{alg:nn-chain}.
Given the forecast ensemble in physical space $\{\mathbf{z}_i^{\mathrm f}\}_{i=1}^{n_e}$ and a prescribed distance metric $\mathsf{D}(\cdot,\cdot)$, the algorithm produces two ordered sequences: a training chain $\textsc{Order} = [i_1, \dots, i_{n_e}]$, which specifies the order in which ensemble members are parameterized through neural-network training, and a parent chain $\textsc{Parent} = [q_1, \dots, q_{n_e}]$, which identifies, for each member, the most similar previously selected ensemble member from which training is initialized.

\begin{algorithm}[!htb]
\caption{Nearest-Neighbor Chain Construction}
\label{alg:nn-chain}
\begin{algorithmic}[1]
\State \textbf{Input:} Forecast ensemble $\{\mathbf{z}_i^{\mathrm f}\}_{i=1}^{n_e}$; distance metric $\mathsf{D}(\cdot,\cdot)$
\State \textbf{Output:} Training chain $\textsc{Order}=[i_1,\dots,i_{n_e}]$; Parent chain $\textsc{Parent}=[q_1,\dots,q_{n_e}]$ with $q_1=-1$
\Statex
\Statex \textbf{Step 1: Identify the most ``central'' ensemble member}
\State Compute pairwise distances $D_{jk} = \mathsf{D}(\mathbf{z}_j^{\mathrm f}, \mathbf{z}_k^{\mathrm f})$ for $1 \le j,k \le n_e$
\State $i_1 \gets \arg\min_{j} \sum_{k=1}^{n_e} D_{jk}$
\State Initialize $\mathcal{A} \gets \{i_1\}$; $\textsc{Order} \gets [i_1]$; $q_1 \gets -1$ \Comment{first network, no parent}
\Statex
\Statex \textbf{Step 2: Construct the nearest-neighbor chain}
\For{$m = 2$ to $n_e$}
    \For{each $r \in \{1,\dots,n_e\} \setminus \mathcal{A}$}
        \State $\hat{a}(r) \gets \arg\min_{a \in \mathcal{A}} D_{ra}$ \Comment{nearest selected neighbor of $r$}
        \State $d(r) \gets D_{r, \hat{a}(r)}$
    \EndFor
    \State $i_m \gets \arg\min_{r} d(r)$ \Comment{next member added to chain}
    \State $q_m \gets \hat{a}(i_m)$ \Comment{parent of $i_m$}
    \State Append $i_m$ to $\textsc{Order}$ and $q_m$ to $\textsc{Parent}$
    \State $\mathcal{A} \gets \mathcal{A} \cup \{i_m\}$ \Comment{update selected set}
\EndFor
\end{algorithmic}
\end{algorithm}

The construction proceeds in two steps. 
In the first step, the most ``central'' ensemble member---defined as the member with the smallest total (or equivalently, average) distance to all other ensemble members---is selected to initialize the chain. Distances are measured using the normalized Euclidean metric
\begin{equation}
\mathsf{D}(\mathbf{z}_j^{\mathrm f}, \mathbf{z}_k^{\mathrm f})
= \sqrt{\frac{1}{n_z}
\sum_{\ell=1}^{n_z}
\left[\mathbf{z}_j^{\mathrm f} - \mathbf{z}_k^{\mathrm f}\right]_\ell^2},
\label{eq:distance}
\end{equation}
where $n_z$ denotes the dimension of the state $\mathbf{z}$ and $[\bm{a}]_\ell$ denotes the $\ell$th component of a vector $\bm{a}$.
This member is then added to the selected subset $\mathcal{A}$ and assigned no parent ($q_1 = -1$).

In the second step, the remaining ensemble members are appended to the chain iteratively based on their similarity to the selected subset. At each iteration, all unselected members are compared to the currently selected set, and the member that is closest to this set under the prescribed distance metric is added next. The nearest previously selected member is recorded as its parent and serves as the source of initialization for training. The selected subset $\mathcal{A}$ is then updated, and this procedure is repeated until all ensemble members are included in the chain.

Once the training chain $\textsc{Order}$ and the corresponding parent chain $\textsc{Parent}$ are established, the neural networks are trained sequentially following $\textsc{Order}$. The first member, corresponding to the most ``central'' member, has no parent and is trained independently from scratch. Each subsequent member is initialized from its parent as specified by $\textsc{Parent}$ and trained with a reduced learning rate to limit parameter drift in the neural space. Training is terminated once a prescribed fitting error threshold is reached, typically on the order of $10^{-6}$. This transfer learning strategy promotes smooth variation of network parameters across the ensemble and yields well-behaved ensemble statistics that support stable EnKF updates.

We note that the nearest-neighbor chain can naturally be generalized to a tree-like structure, in which multiple ensemble members share the same parent ensemble member. In this setting, once a parent ensemble member has been trained, its child ensemble members can be trained in parallel, thereby further improving training efficiency. We leave a systematic optimization of the transfer-learning process for future work and focus here on how the neural EnKF can overcome difficulties in DA for shock-laden flows.

\subsection{Revisiting the motivation example}

\begin{figure}[!htb]
\centering
\includegraphics[width=0.99\textwidth]{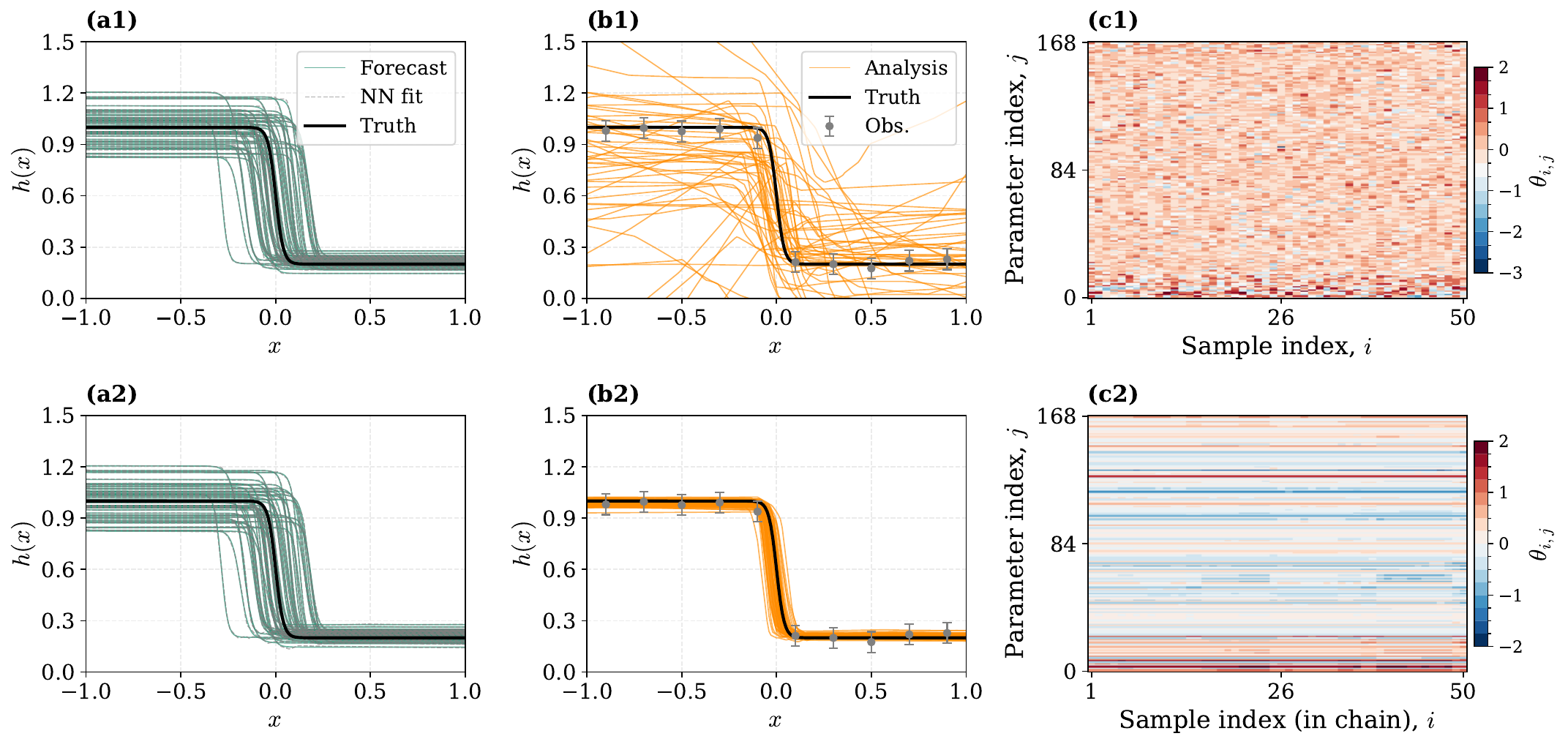}
    \caption{Comparison of neural EnKF updates under two training strategies for the hyperbolic-tangent surrogate shock example. Panels (a1--c1) correspond to \emph{independent training with random initialization}, while panels (a2--c2) correspond to the proposed \emph{nearest-neighbor chain training}.
    (a1) Forecast ensemble together with the corresponding neural-network fits for each ensemble member, indicating negligible fitting errors.
    (b1) Analysis ensemble obtained using independent training, exhibiting irregular and spatially incoherent updates across ensemble members.
    (c1) Variation of neural-network parameters across ensemble members under independent training, showing highly irregular variability and poor alignment of parameters in neural space.
    (a2--b2) Forecast and analysis ensembles obtained using nearest-neighbor chain training, in which the analysis ensemble preserves the shock-like structure and better reflects the observation data.
    (c2) Smooth neural-network parameter variations under nearest-neighbor chain training, showing well-aligned parameter distributions in neural space.
    In panels (c1) and (c2), the horizontal axis indexes the ensemble members (50 samples), while the vertical axis indexes the flattened neural-network parameters (169 trainable parameters).}
    \label{fig:tanh-neuralEnKF}
\end{figure}

We revisit the hyperbolic-tangent surrogate shock example from Section~\ref{sec:motivation} to assess the effectiveness of the neural EnKF with nearest-neighbor chain training. We compare the neural EnKF updates under two training strategies: independent training with random initialization and the proposed nearest-neighbor chain training, as illustrated in Fig.~\ref{fig:tanh-neuralEnKF}. The comparison demonstrates that nearest-neighbor chain training enables a stable, structure-preserving neural EnKF update, whereas independent training with random initialization does not.

As shown in panels (a1) and (a2), both training strategies provide accurate neural-network representations for the forecast ensemble members. Under independent training with random initialization, however, the subsequent neural EnKF update produces a severely distorted analysis ensemble (b1). These distortions are not confined to the shock region but extend across the entire domain, including smooth regions where no sharp features are present. This behavior reflects a failure of the neural-space update and, in this example, results in analysis states that are even worse than those obtained using the EnKF (see Fig.~\ref{fig:tanh-EnKF}(b)).
The origin of this failure is evident in the neural-network parameter variations shown in panel (c1). Under independent training, the parameter values across ensemble members exhibit highly irregular and non-smooth variations, indicating poor alignment in neural space, which in turn leads to ineffective neural EnKF updates.
In contrast, panel (c2) exhibits smooth parameter variations across ensemble members under nearest-neighbor chain training. Most neural-network parameters take nearly identical values across the ensemble, with only small and gradually varying deviations along the chain, indicating a highly concentrated and well-aligned parameter distribution in neural space. This alignment leads to effective neural EnKF updates, yielding a structure-preserving analysis ensemble shown in panel (b2) without spurious oscillations.

\section{Numerical experiments with neural EnKF}
\label{sec:result}

The neural EnKF is evaluated on three test cases of increasing complexity and consistently demonstrates robust DA performance, effectively preserving sharp flow structures while avoiding spurious oscillations.
The test suite consists of three representative problems:
(1) the inviscid Burgers’ equation;
(2) the Sod shock tube; and
(3) a 2D blast wave.

\subsection{Numerical experiments on the inviscid Burgers' equation}
\begin{figure}[!htb]
\centering
\includegraphics[width=0.99\textwidth]{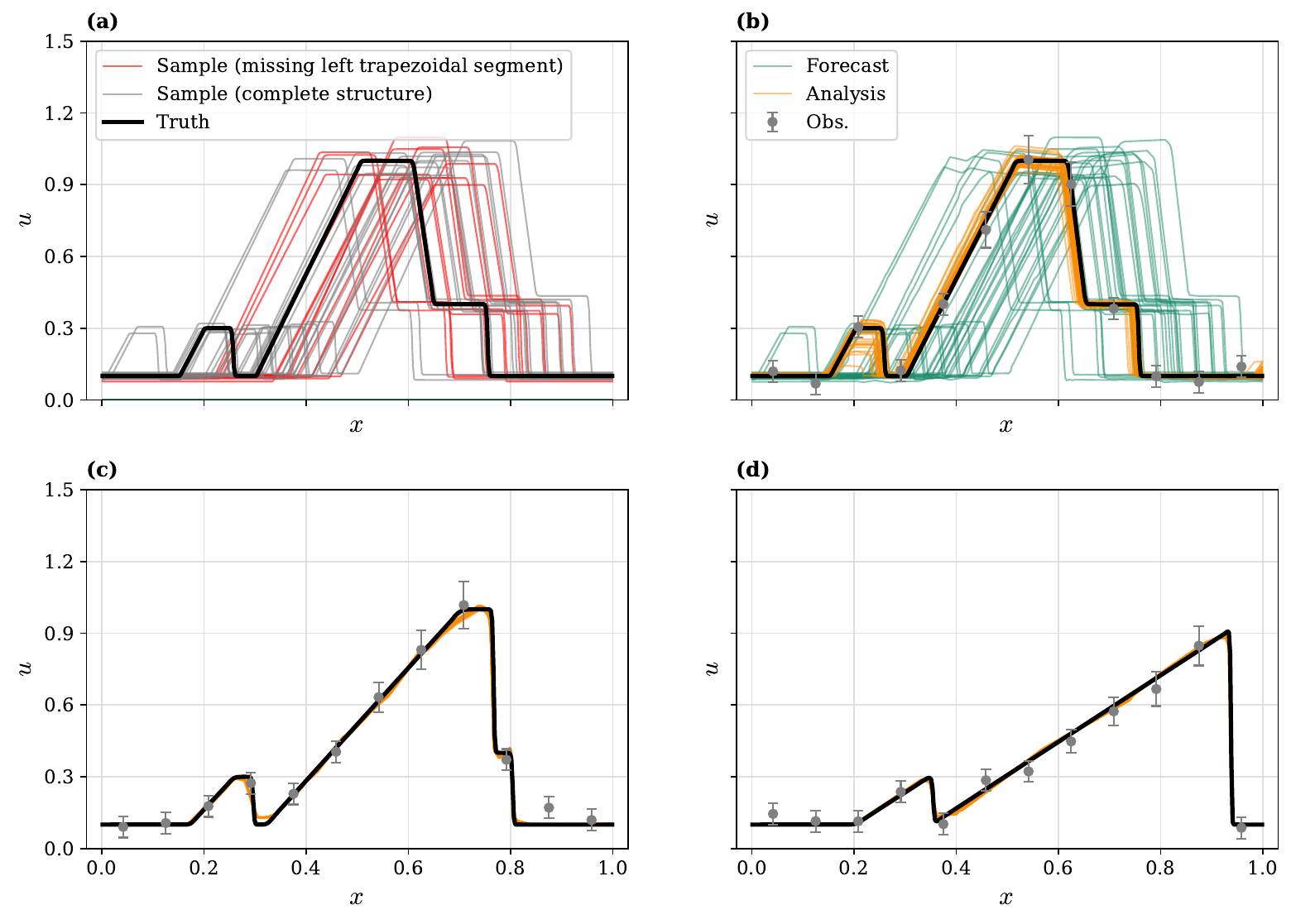}
    \caption{DA results for the inviscid Burgers’ equation with a structurally heterogeneous initial ensemble under neural EnKF updates. 
    (a) Initial ensemble, where several members (red) lack the left trapezoidal segment present in the true profile (black), while the remaining members (grey) retain the complete structure. 
    (b) Forecast and analysis ensembles at the first DA cycle, illustrating rapid recovery of the overall profile. 
    (c) and (d) Analysis ensembles at the 20th and 50th DA cycles, respectively, showing progressive structural convergence toward the true solution during intermediate and late stages of assimilation.}
    \label{fig:da-trapezoidal}
\end{figure}

We consider the inviscid Burgers’ equation
\begin{equation}
\frac{\partial u}{\partial t}
+ u \frac{\partial u}{\partial x}
= 0,
\end{equation}
on the domain $x \in [0,1]$ with periodic boundary conditions, where $u$ denotes the scalar velocity field. The reference initial velocity profile is piecewise linear and contains multiple sharp features, including a small trapezoidal segment on the left (Fig.~\ref{fig:da-trapezoidal}(a)).

The domain is discretized using 400 uniformly spaced grid points. The equation is solved using the \texttt{PyClaw} library~\cite{pyclaw-sisc} and advanced to a final time $T = 0.5$ with a fixed time step of $\Delta t = 2 \times 10^{-4}$.

To model uncertainty in the initial condition, we construct an ensemble of $n_e = 40$ members. While the reference initial profile contains the left trapezoidal feature, several ensemble members omit this structure, resulting in pronounced structural heterogeneity across the ensemble (Fig.~\ref{fig:da-trapezoidal}(a), where the members missing the left trapezoidal segment are highlighted in red). Consequently, the ensemble does not share a consistent representation of the underlying solution features. This configuration provides a stringent test for DA methods, as it violates the structural similarity assumptions typically required by feature alignment-based approaches (e.g.,~\cite{subrahmanya2025feature}).

The large initial uncertainty is progressively reduced through sequential DA using sparse, noisy observations. Neural EnKF updates are applied at regular intervals of $\Delta t_{\mathrm{obs}} = 0.01$ starting at $t = 0.01$, resulting in 50 DA cycles.
At each DA step, each forecast ensemble member $\mathbf{z}_i^{\mathrm{f}} = \bm{u}_i^{\mathrm{f}}$ is represented by a neural network with parameters $\bm{\theta}_i^{\mathrm{f}}$, with the network architecture detailed in~\ref{app:nn-arch}.
Synthetic observations $\mathbf{d}$ are generated by sampling the true velocity field $\bm{u}^*$ at 12 uniformly distributed interior locations over the domain, i.e., $x_k = 1/24 + k/12$ for $k = 0,\dots,11$. 
The resulting values are perturbed by additive Gaussian noise,
\begin{equation}
\mathbf{d} = \mathsf{H}\bm{u}^* + \bm{\eta},
\qquad
\bm{\eta} \sim \mathcal{N}(\mathbf{0}, \mathbf{R}),
\label{eq:burgers-obs}
\end{equation}
where $\mathbf{R}$ is a diagonal covariance matrix with entries $R_{kk} = \bigl(\max(0.05\,[\mathsf{H}\bm{u}^*]_k,\, 5\times10^{-3})\bigr)^2$.
Consistent with the observation model, the predicted observations for each forecast ensemble member are
\begin{equation}
\mathbf{y}_i = \mathsf{H}\bm{u}_i^\mathrm{f}.
\label{eq:burgers-predict-obs}
\end{equation}
Given the ensemble of network parameters $\{\bm{\theta}_i^{\mathrm{f}}\}_{i=1}^{40}$, the corresponding predicted observations $\{\mathbf{y}_i\}_{i=1}^{40}$, and the synthetic observations $\mathbf{d}$, the neural EnKF update is performed according to Eq.~\eqref{eq:neural-enkf-update}. 
The updated network parameters are then mapped back to physical space to obtain the analysis ensemble, as described in Eq.~\eqref{eq:reconstruction}.

The neural EnKF exhibits stable and effective DA performance in this challenging setting, with rapid convergence toward the true solution, as shown in Fig.~\ref{fig:da-trapezoidal}. After the first DA cycle, the analysis ensemble shows a marked reduction in spread and improved agreement with the true profile over most of the domain, including partial reconstruction of the missing left trapezoidal feature in the members that initially lack this structure. By the 20th DA cycle, the ensemble continues to approach the true solution, with these members progressively recovering the absent feature. By the 50th DA cycle, the analysis ensemble closely matches the true velocity profile across the entire domain, indicating robust recovery despite the substantial initial structural mismatch.

\begin{figure}[!htb]
\centering
\includegraphics[width=0.55\textwidth]{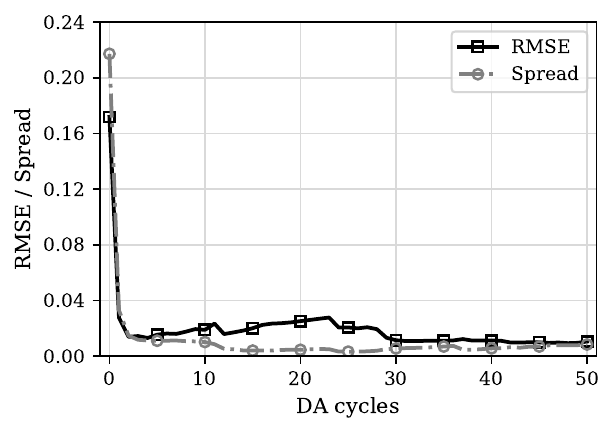}
    \caption{Evolution of RMSE and ensemble spread over DA cycles for the inviscid Burgers’ equation with an initial ensemble exhibiting structural heterogeneity. Both metrics decrease rapidly during the initial assimilation stage and subsequently stabilize at comparable low levels. Markers are shown every five DA cycles for clarity.}
    \label{fig:rmse-trapezoidal}
\end{figure}

The DA performance is further quantified by examining the evolution of the root-mean-square error (RMSE) and the ensemble spread over successive DA cycles. The RMSE is defined as
\begin{equation}
\mathrm{RMSE}
=
\sqrt{
\frac{1}{n_z}
\sum_{\ell=1}^{n_z}
\left[ \bar{\mathbf{z}}^{\mathrm{a}} - \mathbf{z}^* \right]_\ell^2
}, \quad \quad
\bar{\mathbf{z}}^{\mathrm{a}}
= \frac{1}{n_e} \sum_{i=1}^{n_e} \mathbf{z}_i^{\mathrm{a}},
\label{eq:ens-rmse}
\end{equation}
where $\bar{\mathbf{z}}^{\mathrm{a}}$ and $\mathbf{z}^*$ denote the analysis ensemble mean and true state at a given DA step. The ensemble spread is defined as
\begin{equation}
\mathrm{Spread}=\sqrt{\frac{1}{n_z} \operatorname{trace}\left(\mathbf{P}^{\mathrm{a}}\right)}, \quad \quad \mathbf{P}^{\mathrm{a}}
=
\frac{1}{n_e-1}
\sum_{i=1}^{n_e}
\left(\mathbf{z}_i^{\mathrm{a}}-\bar{\mathbf{z}}^{\mathrm{a}}\right)
\left(\mathbf{z}_i^{\mathrm{a}}-\bar{\mathbf{z}}^{\mathrm{a}}\right)^{\top}.
\label{eq:ens-spread}
\end{equation}
As shown in Fig.~\ref{fig:rmse-trapezoidal}, both the RMSE and the spread decrease rapidly during the initial DA cycles and subsequently stabilize at comparably low levels.
Importantly, the close agreement between the RMSE and the spread at later DA stages (i.e., $\mathrm{RMSE} \approx \text{spread}$) indicates 
consistency between the estimation error and the ensemble uncertainty.
We note that here the analysis ensemble mean is computed by averaging the reconstructed ensemble members in physical space. Alternatively, one may first compute the ensemble mean in the neural space and then reconstruct the corresponding physical state, which may yield sharper shock features in the ensemble mean.

\subsection{Numerical experiments on the Sod shock tube problem}
We consider the Sod shock tube problem~\cite{sod1978survey}, governed by the compressible Euler equations in conservative form,  
\begin{equation}
\frac{\partial \bm{W}}{\partial t}
+ \nabla \cdot \mathcal{F}(\bm{W}) = 0,
\label{eq:euler}
\end{equation}
where $\bm{W}$ denotes the conservative state vector and $\mathcal{F}(\bm{W})$ denotes the corresponding advective fluxes.  
These are defined as
\begin{equation}
\bm{W} =
\left[\begin{array}{c}
\rho \\
\rho \bm{V} \\
\rho E
\end{array}\right],
\qquad
\mathcal{F}(\bm{W}) =
\left[\begin{array}{c}
\rho \bm{V}^{\top} \\
\rho \bm{V} \otimes \bm{V} + p \bm{I} \\
(\rho E + p)\bm{V}^{\top}
\end{array}\right],
\label{eq:flux}
\end{equation}
where $\rho$ and $p$ are the density and pressure, $\bm{V} = [u, v, w]^{\top}$ is the velocity vector, $\bm{I}$ is the identity matrix, and the operator $\otimes$ denotes the outer product.
The total energy per unit mass is 
\begin{equation}
E = e + \frac{1}{2} |\bm{V}|^2,
\label{eq:total_energy}
\end{equation}
where $e$ is the internal energy per unit mass and $|\bm{V}|$ denotes the Euclidean norm of the velocity vector.  
The system is closed by the ideal-gas equation of state,  
\begin{equation}
p = (\gamma - 1)\rho e,
\label{eq:eos}
\end{equation}
with $\gamma = 1.4$. In the 1D setting considered here, the velocity reduces to $\bm{V} = [u, 0, 0]^{\top}$.

\begin{figure}[!htb]
\centering
\includegraphics[width=0.66\textwidth]{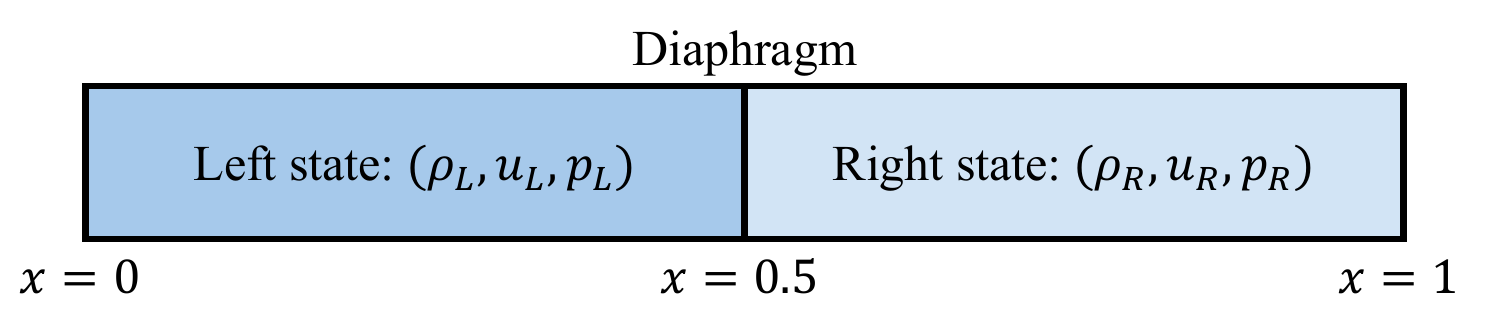}
\caption{Schematic of the reference shock tube configuration. The diaphragm at $x_d = 0.5$ separates the high-pressure left state $(\rho_L, u_L, p_L)$ and the low-pressure right state $(\rho_R, u_R, p_R)$.}
\label{fig:sod-schematic}
\end{figure}

The problem is considered on the domain $x \in [0,1]$, with non-reflecting boundary conditions imposed at both ends to permit outgoing waves to exit without artificial reflection. The reference initial condition consists of a diaphragm located at $x_d = 0.5$ separating two constant states,
\begin{equation}
(\rho_L, u_L, p_L) = (1.0,\, 0,\, 1.0), 
\qquad
(\rho_R, u_R, p_R) = (0.125,\, 0,\, 0.1),
\label{eq:sod_ic}
\end{equation}
corresponding to high- and low-pressure regions, respectively, as illustrated in Fig.~\ref{fig:sod-schematic}.

The governing equations are discretized on a uniform grid of 400 cells and solved using the open-source solver \texttt{M2C} (Multiphysics Modeling and Computation)~\cite{zhao2026m2c,wang_m2c}. Numerical fluxes are computed using a Roe-type approximate Riemann solver. Time integration is performed using an explicit second-order Runge--Kutta scheme with a fixed time step of $\Delta t = 5 \times 10^{-4}$, corresponding to a Courant--Friedrichs--Lewy (CFL) number of approximately 0.4. The simulation is advanced to a final time of $T = 0.2$, a standard configuration for this benchmark problem.

\begin{figure}[!htb]
\centering
\includegraphics[width=0.99\textwidth]{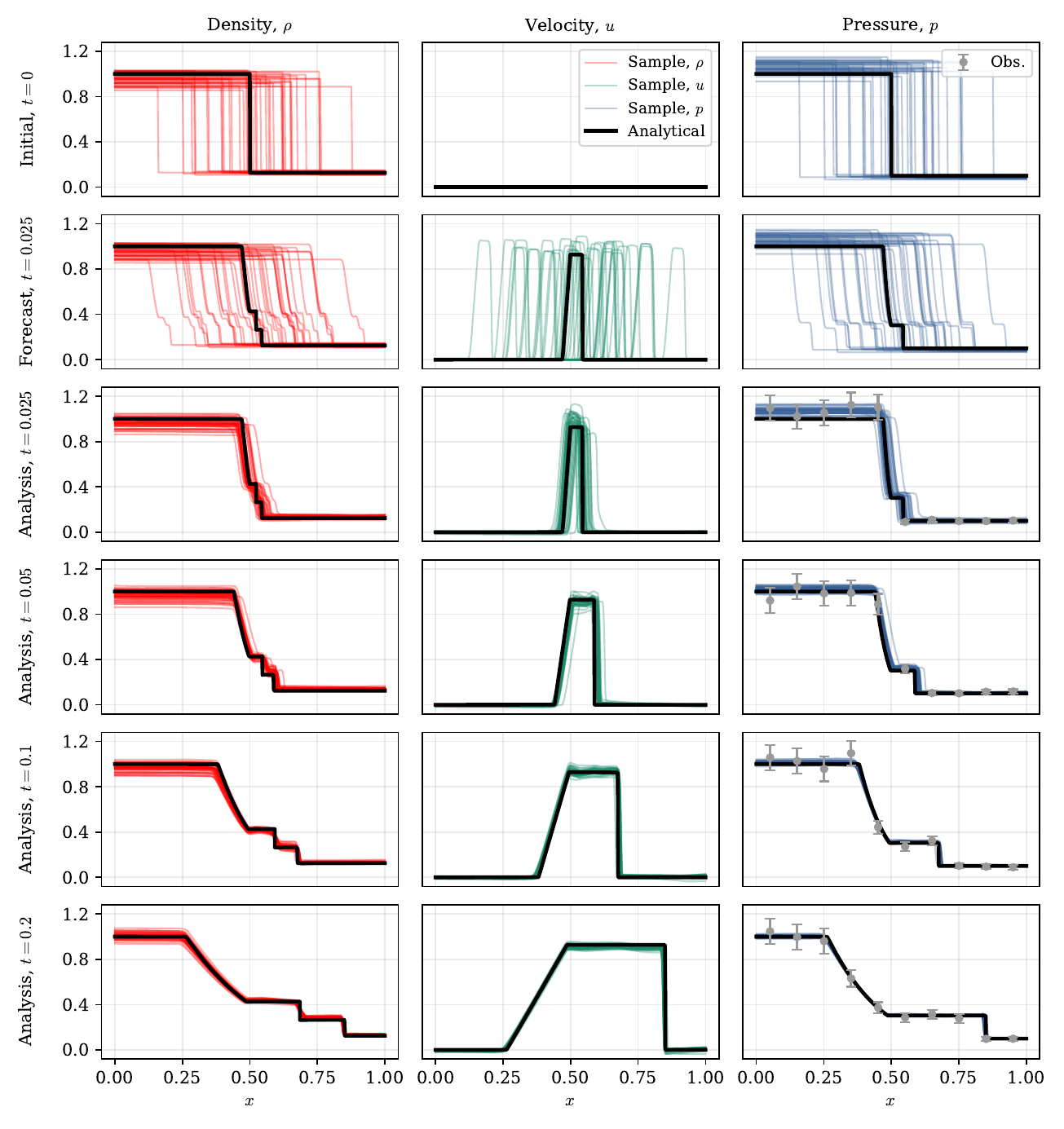}
\caption{Analysis ensemble for the Sod shock tube problem at representative DA steps, showing increasing agreement with the analytical solution. Columns show density $\rho$, velocity $u$, and pressure $p$. Rows correspond to the initial ensemble ($t=0$), the forecast ensemble prior to the first DA step ($t=0.025$), and the analysis ensembles at the 1st, 2nd, 4th, and 8th DA steps ($t = 0.025$, $0.05$, $0.1$, and $0.2$). The density field exhibits three flow features, from left to right: a rarefaction wave, a contact interface, and a shock, whereas the velocity and pressure fields contain a rarefaction wave and a shock but do not include a contact discontinuity.}
\label{fig:sod-analysis}
\end{figure}

To represent uncertainty in the initial condition, we construct an ensemble of $n_e = 40$ perturbed states by independently sampling the left- and right-state parameters, as well as the diaphragm location, from truncated Gaussian distributions:
\begin{equation}
\begin{aligned}
\rho_L &\sim \mathcal{N}_{(0,\infty)}\!\left(0.95,\,0.05^{2}\right), \qquad
p_L \sim \mathcal{N}_{(0,\infty)}\!\left(1.05,\,0.05^{2}\right), \qquad
x_d \sim \mathcal{N}_{(0.1, 0.9)}\!\left(0.55,\,0.2^{2}\right), \\
\rho_R &\sim \mathcal{N}_{(0,\infty)}\!\left(0.13,\,0.01^{2}\right), \qquad
p_R \sim \mathcal{N}_{(0,\infty)}\!\left(0.09,\,0.01^{2}\right).
\end{aligned}
\label{eq:sod-sampling}
\end{equation}
Each sampled parameter set defines one ensemble member, as illustrated in the top row of Fig.~\ref{fig:sod-analysis}.
We note that the means of the sampling distributions are intentionally biased relative to the reference values, a setting that may better reflect practical DA scenarios and was not examined in some related studies (e.g.,~\cite{subrahmanya2025feature}).

The large initial uncertainty is progressively reduced through sequential DA using the neural EnKF with sparse, noisy pressure observations. Neural EnKF updates are applied every $\Delta t_{\mathrm{obs}} = 0.025$ starting at $t = 0.025$, yielding eight DA steps.
At each DA step, each forecast ensemble member $\mathbf{z}_i^{\mathrm{f}} = [\bm{\rho}_i^{\mathrm{f}}, \bm{u}_i^{\mathrm{f}}, \bm{p}_i^{\mathrm{f}}]$ represents the full state vector, consisting of the density, velocity, and pressure components.
The entire state is parameterized by a single neural network with parameters $\bm{\theta}_i^{\mathrm{f}}$, which implicitly represents the coupled flow variables.
Synthetic pressure observations $\mathbf{d}$ are constructed by sampling the analytical pressure solution $\bm{p}^*$ (evaluated on the same spatial grid as the numerical solution) at 10 uniformly spaced locations $x = 0.05, 0.15, \ldots, 0.95$, and perturbing the sampled values with additive Gaussian noise:
\begin{equation}
    \mathbf{d} = \mathsf{H}\bm{p}^* + \bm{\eta},
    \qquad
    \bm{\eta} \sim \mathcal{N}(\mathbf{0}, \mathbf{R}),
    \label{eq:sod-obs}
\end{equation}
where $\mathbf{R}$ is a diagonal covariance matrix with entries $R_{kk} = \left(0.05\,[\mathsf{H}\bm{p}^*]_k + 0.01\right)^2$.
We note that the analytical solution is used to generate the synthetic observations, whereas forward simulations are performed using a finite-resolution numerical solver. This deliberate model–data mismatch introduces a more realistic and challenging DA setting.
Consistent with the observation model, the predicted observations associated with each forecast ensemble member are given by
\begin{equation}
    \mathbf{y}_i = \mathsf{H}\bm{p}_i^\mathrm{f}.
    \label{eq:sod-predict-obs}
\end{equation}
Given the ensemble of neural-network parameters $\{\bm{\theta}_i^{\mathrm{f}}\}_{i=1}^{40}$, the corresponding predicted observations $\{\mathbf{y}_i\}_{i=1}^{40}$, and the synthetic pressure observations $\mathbf{d}$, the neural EnKF update is performed according to Eq.~\eqref{eq:neural-enkf-update}.
The updated network parameters are then mapped back to physical space through the reconstruction described in Eq.~\eqref{eq:reconstruction}.

The neural EnKF demonstrates robust DA performance, progressively recovering the true solution over successive assimilation steps without introducing spurious oscillations. 
For comparison, results obtained with the EnKF are reported in~\ref{app:sod-enkf}, where strong oscillations and nonphysical states (negative density and pressure) are observed.
Fig.~\ref{fig:sod-analysis} illustrates the behavior of the neural EnKF through the evolution of the ensemble across successive assimilation steps. Prior to the first DA update ($t = 0.025$), the forecast ensemble exhibits substantial spread and pronounced misalignment of discontinuities, reflecting significant uncertainty in the locations of the shock, contact interface, and rarefaction wave.
Following the first DA step, the analysis ensemble contracts noticeably and exhibits improved alignment of the dominant flow features. With successive DA updates, the ensemble progressively sharpens around the analytical solution, and the spread associated with misaligned discontinuities is systematically reduced. By the later assimilation times, the ensemble members are tightly clustered and the recovered solution accurately captures the rarefaction wave, contact interface, and shock without introducing spurious oscillations.

The convergence behavior, however, differs across state variables. 
The pressure field, which is directly observed, converges most rapidly, while the velocity field exhibits a comparable convergence behavior due to its strong dynamical coupling with pressure through the momentum equation. 
In contrast, the density field shows a more gradual reduction of ensemble uncertainty: although the analysis ensemble progressively recovers the contact interface and overall solution structure, 
the contraction of spread lags behind that of pressure and velocity. 
This behavior is expected, since pressure observations alone do not uniquely determine the density field through the equation of state without additional thermodynamic information, such as temperature or internal energy. 
As a result, multiple density--energy configurations can remain consistent with the assimilated pressure observations, 
limiting the contraction of density uncertainty and leading to a slower reduction of spread.
A more detailed analysis of this effect is provided in~\ref{app:sod-sensitivity}.

\begin{figure}[!htb]
\centering
\includegraphics[width=0.99\textwidth]{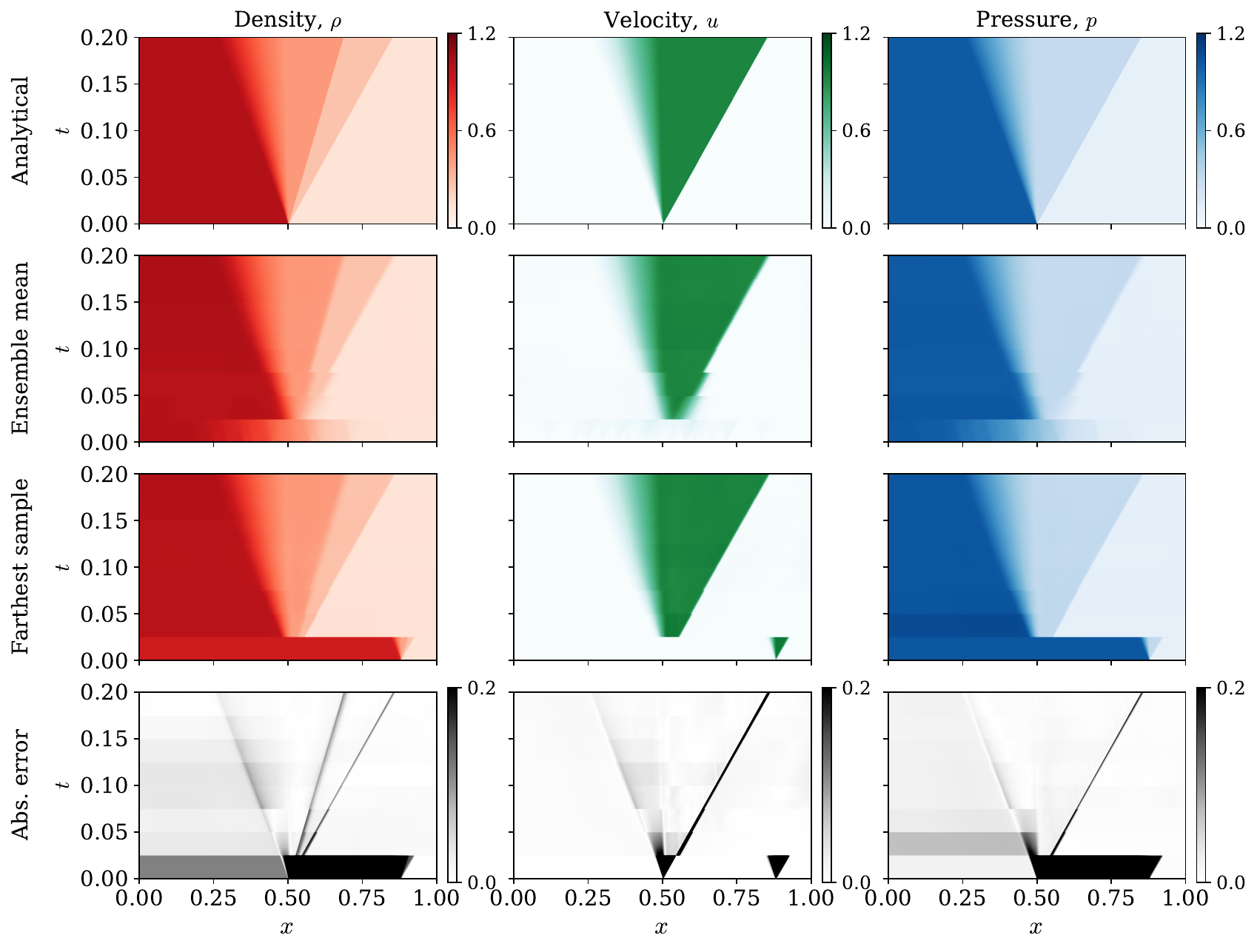}
\caption{Comparison of the space--time evolution under DA for the Sod shock tube problem, with neural EnKF updates applied every $\Delta t_{\mathrm{obs}} = 0.025$. Columns show density $\rho$, velocity $u$, and pressure $p$. Rows correspond to the analytical solution, ensemble mean, a ``farthest'' ensemble member, and the absolute error between the analytical solution and the farthest ensemble member. The farthest ensemble member, defined once as the last member in the nearest-neighbor chain at the first DA step, corresponds to the member with the largest initial structural mismatch. The analytical solution and error fields are plotted on a fine grid with 3200 cells to enhance visualization of sharp discontinuities, while the ensemble mean and farthest ensemble member are shown on the numerical grid with 400 cells.}
\label{fig:sod-contour}
\end{figure}

To further illustrate the dynamical behavior under DA, Fig.~\ref{fig:sod-contour} compares the space--time evolution of the analytical solution with that of the ensemble mean and a ``farthest'' ensemble member over the full simulation window.
Here, the ensemble mean represents a statistical state estimate rather than a physical solution. 
The farthest ensemble member is defined once as the last member in the nearest-neighbor chain constructed at the first DA step, corresponding to the ensemble member with the largest initial structural mismatch. This ensemble member is then tracked throughout the assimilation process and evolves according to the governing equations between successive DA updates.
Since neural EnKF updates are applied at regular intervals of $\Delta t_{\mathrm{obs}} = 0.025$, both the ensemble-mean and farthest-member evolutions exhibit a piecewise-in-time structure, with each segment corresponding to forward integration between DA steps. As DA proceeds, the ensemble mean (second row) becomes progressively better aligned with the analytical solution (top row), reflecting the improvement of the ensemble-averaged state estimate. The evolution of the farthest ensemble member (third row) provides complementary insight into the behavior of individual realizations under DA. Despite its unfavorable initial condition, this trajectory rapidly adjusts at the first DA step ($t = 0.025$) and subsequently tracks the analytical solution with increasing fidelity, indicating that the neural EnKF effectively corrects pronounced structural misalignment of dominant flow features even for extreme ensemble members.
These trends are further highlighted by the absolute-error fields (bottom row), which show the pointwise differences between the analytical solution and the farthest ensemble member. While noticeable discrepancies are present near discontinuities at early times, they decrease systematically over successive DA cycles. By the final DA step, the remaining differences are confined to narrow regions around the shock and contact interface and are primarily attributable to the finite spatial resolution of the numerical solver rather than deficiencies of the neural EnKF.

\begin{figure}[!htb]
\centering
\includegraphics[width=0.99\textwidth]{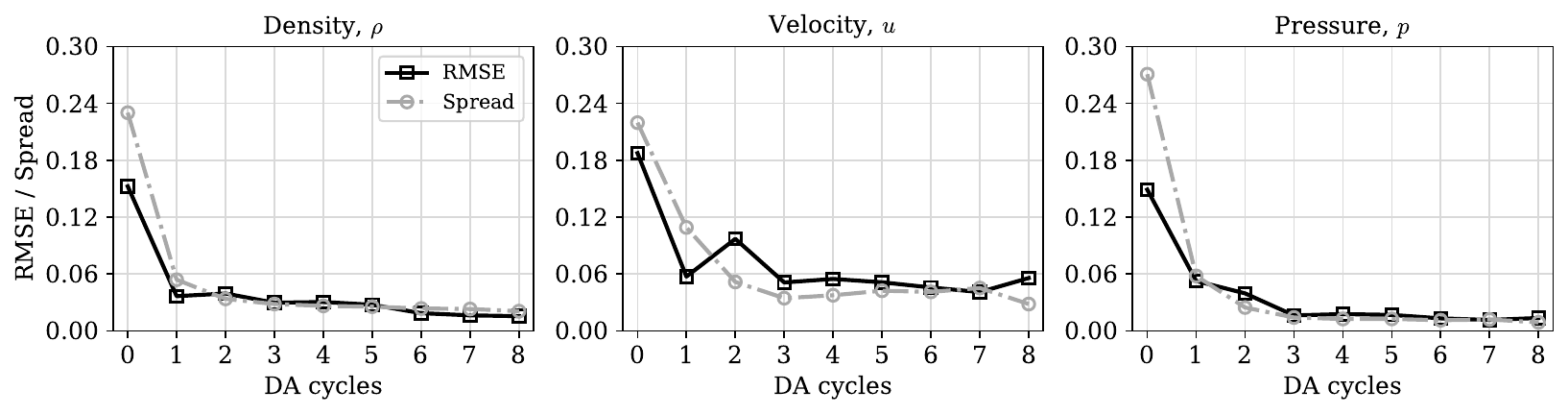}
\caption{RMSE and ensemble spread over successive DA cycles for the Sod shock tube problem under the neural EnKF, showing rapid reduction of both metrics across all three variables during the first few DA cycles and sustained comparably low levels thereafter. The value at cycle~0 is computed from the forecast ensemble prior to the first DA update ($t=0.025$), while cycles~1--8 correspond to the analysis ensembles after successive DA steps.}
\label{fig:sod-rmse-spread}
\end{figure}

The DA performance is further quantified using the RMSE and ensemble spread, as shown in Fig.~\ref{fig:sod-rmse-spread}.
Here, the value at cycle~0 is computed from the forecast ensemble prior to the first DA update ($t = 0.025$), while cycles~1--8 correspond to the analysis ensembles after successive DA steps.
For all three state variables, both metrics decrease rapidly during the initial DA cycles, reflecting effective reduction of estimation error and ensemble uncertainty through assimilation.
Following the initial spin-up phase, the RMSE and ensemble spread remain at comparably low levels across subsequent DA cycles, indicating that the state estimate remains reasonably accurate and that ensemble uncertainty is effectively controlled without collapse.

To further evaluate the robustness of the neural EnKF under various DA configurations, we perform a parametric study examining the effects of observation noise level, spatial density of observations, and observation frequency. Detailed experimental setups and results are provided in~\ref{app:sod-parametric}.

\subsection{Numerical experiments on the blast wave problem}

We consider a 2D blast wave problem governed by the same compressible Euler equations as in the shock tube case, i.e., Eqs.~\eqref{eq:euler}–\eqref{eq:eos}. In this 2D setting, the velocity field takes the form $\bm{V} = [u, v, 0]^{\top}$.

\begin{figure}[!htb]
\centering
\includegraphics[width=0.6\textwidth]{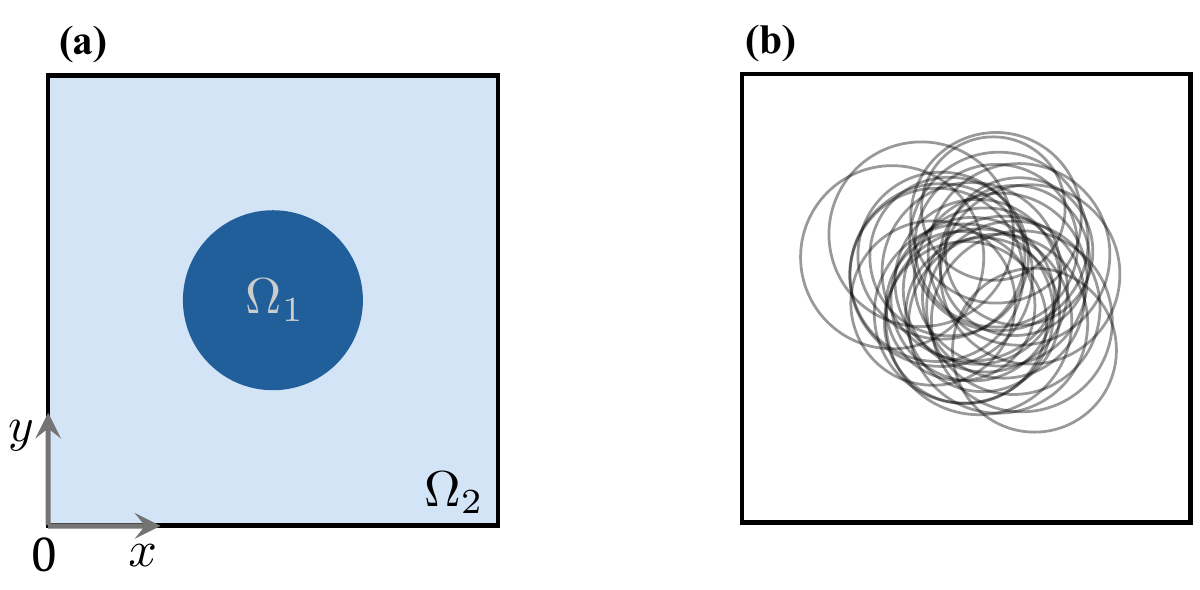}
\caption{Schematic illustration of the 2D blast-wave initial configuration and ensemble sampling. (a) Computational domain $\Omega = \Omega_1 \cup \Omega_2$, where $\Omega_1$ denotes the circular high-pressure blast region and $\Omega_2$ the surrounding low-pressure region. (b) Ensemble of sampled initial discontinuity locations obtained by Gaussian sampling of the blast center and radius. For clarity, uncertainty in the density and pressure within $\Omega_1$ is not shown.}
\label{fig:bw-schematic}
\end{figure}

The problem is considered on the domain $\Omega = [0,1] \times [0,1]$, with non-reflecting boundary conditions imposed on all boundaries to permit outgoing waves to exit without artificial reflection. The initial condition consists of a circular discontinuity separating two constant states, as illustrated in Fig.~\ref{fig:bw-schematic}(a). Specifically, a high-pressure circular region $\Omega_1$ is embedded in a low-pressure exterior region $\Omega_2 = \Omega \setminus \Omega_1$. The reference initial condition is prescribed as
\begin{equation}
(\rho_{\Omega_1}, u_{\Omega_1}, v_{\Omega_1}, p_{\Omega_1}) = (1.0, 0.0, 0.0, 100),
\qquad
(\rho_{\Omega_2}, u_{\Omega_2}, v_{\Omega_2}, p_{\Omega_2}) = (1.0, 0.0, 0.0, 0.1),
\label{eq:bw_ic}
\end{equation}
where $\Omega_1$ is a circular region centered at $(x_c, y_c) = (0.5, 0.5)$ with radius $r = 0.2$. This configuration generates a strong radially propagating blast wave.

The Euler equations are discretized on a uniform Cartesian grid of $128 \times 128$ cells. Time integration is performed using an explicit second-order Runge--Kutta scheme with a fixed time step of $\Delta t = 10^{-4}$, corresponding to a CFL number of approximately 0.2. The simulation is advanced to a final time of $T = 2.0 \times 10^{-2}$.

To represent uncertainty in the initial condition, we construct an ensemble of $n_e = 30$ perturbed states by independently sampling the parameters defining the blast wave configuration from truncated Gaussian distributions. Specifically, the blast center coordinates, radius, density, and pressure are sampled as:
\begin{equation}
\begin{aligned}
x_c &\sim \mathcal{N}_{(0.3, 0.7)}(0.53,\, 0.1^2), \quad
y_c \sim \mathcal{N}_{(0.3, 0.7)}(0.53,\, 0.1^2), \quad
r \sim \mathcal{N}_{(0.16, 0.24)}(0.19,\, 0.02^2), \\
\rho_{\Omega_1} &\sim \mathcal{N}_{(0, \infty)}(1.05,\, 0.1^2), \quad
p_{\Omega_1} \sim \mathcal{N}_{(0, \infty)}(102.0,\, 3.0^2).
\end{aligned}
\label{eq:bw-sampling}
\end{equation}
Each sampled parameter set defines one ensemble member, resulting in an ensemble of initial blast configurations. The induced variability in the initial discontinuity location is illustrated in Fig.~\ref{fig:bw-schematic}(b).
As in the Sod shock tube case, the means of the sampling distributions are intentionally biased relative to the corresponding reference values.

The resulting uncertainty is progressively reduced through sequential DA using the neural EnKF with sparse, noisy pressure observations. Neural EnKF updates are applied every $\Delta t_{\mathrm{obs}} = 4.0 \times 10^{-3}$ starting at $t = 4.0 \times 10^{-3}$, yielding five DA steps.
At each DA step, each forecast ensemble member
$\mathbf{z}_i^{\mathrm{f}} = [\bm{\rho}_i^{\mathrm{f}}, \bm{u}_i^{\mathrm{f}}, \bm{v}_i^{\mathrm{f}}, \bm{p}_i^{\mathrm{f}}]$
represents the full state vector, comprising the density, velocity components in the $x$- and $y$-directions, and pressure. The entire state is parameterized by a single neural network with parameters $\bm{\theta}_i^{\mathrm{f}}$.
Synthetic pressure observations $\mathbf{d}$ are constructed by sampling the reference pressure field $\bm{p}^*$ on a uniform $7 \times 7$ grid of observation locations, with coordinates
$(x,y) \in \{0.05, 0.2, \ldots, 0.95\} \times \{0.05, 0.2, \ldots, 0.95\}$,
and adding independent Gaussian noise:
\begin{equation}
    \mathbf{d} = \mathsf{H}\bm{p}^* + \bm{\eta},
    \qquad
    \bm{\eta} \sim \mathcal{N}(\mathbf{0}, \mathbf{R}).
\end{equation}
where $\mathbf{R}$ is a diagonal covariance matrix whose entries are given by
$R_{kk} = \left(0.03\,[\mathsf{H}\bm{p}^*]_k + 10^{-3}\right)^2$.
Consistent with the observation model, the predicted observations for each forecast ensemble member are
\begin{equation}
    \mathbf{y}_i = \mathsf{H}\bm{p}_i^{\mathrm{f}}.
\end{equation}
Given the ensemble of neural-network parameters $\{\bm{\theta}_i^{\mathrm{f}}\}_{i=1}^{30}$, the corresponding predicted observations $\{\mathbf{y}_i\}_{i=1}^{30}$, and the synthetic observation $\mathbf{d}$, the neural EnKF update is performed according to Eq.~\eqref{eq:neural-enkf-update}, after which the physical-space analysis ensemble is reconstructed following Eq.~\eqref{eq:reconstruction}.

\begin{figure}[!htb]
\centering
\includegraphics[width=0.99\textwidth]{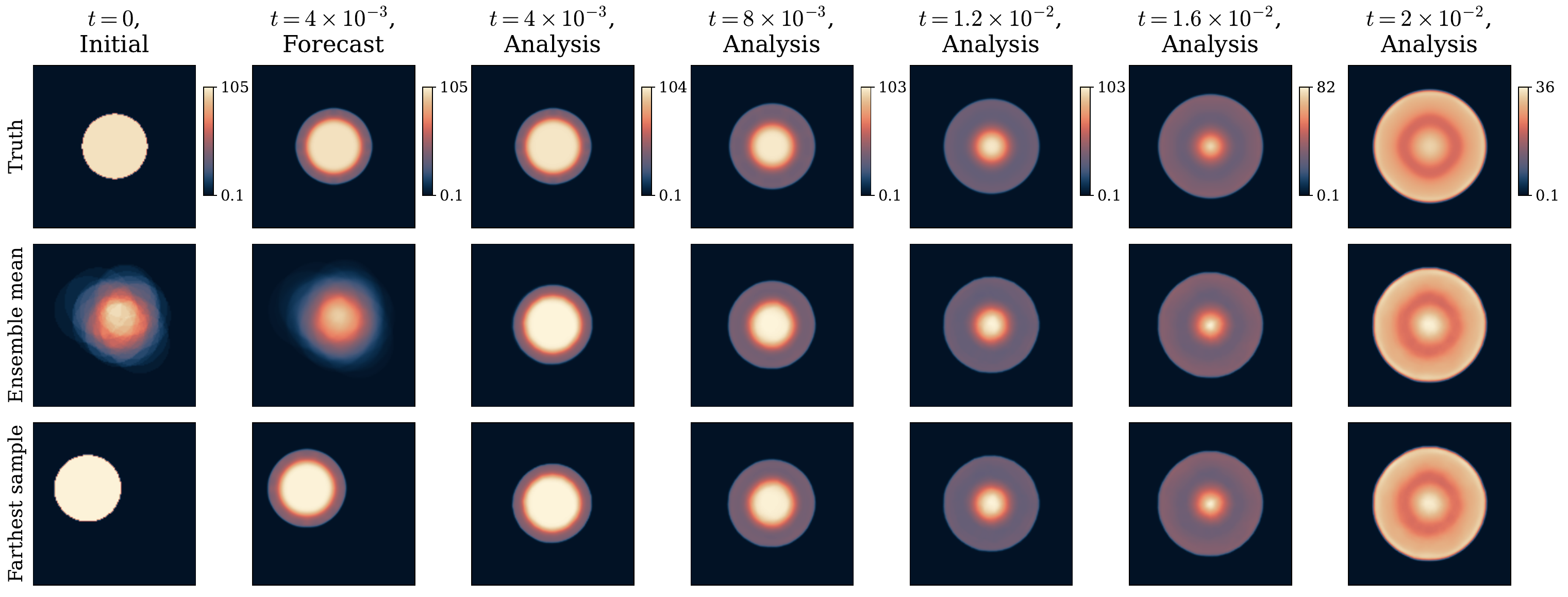}
\caption{
Pressure fields for the 2D blast wave over the full assimilation window, demonstrating rapid correction after the first neural EnKF update and sustained agreement with the reference solution thereafter. 
Rows show the reference solution, the ensemble mean (statistical state estimate), and the farthest ensemble member, defined as the last member in the nearest-neighbor chain constructed at the first DA step. 
Columns correspond to the initial condition ($t=0$), the forecast states prior to the first DA update ($t = 4.0\times10^{-3}$), and the analysis states after five successive DA steps up to the final time ($t = 2.0\times10^{-2}$).
}
\label{fig:bw_p_analysis}
\end{figure}

\begin{figure}[!htb]
\centering
\includegraphics[width=0.99\textwidth]{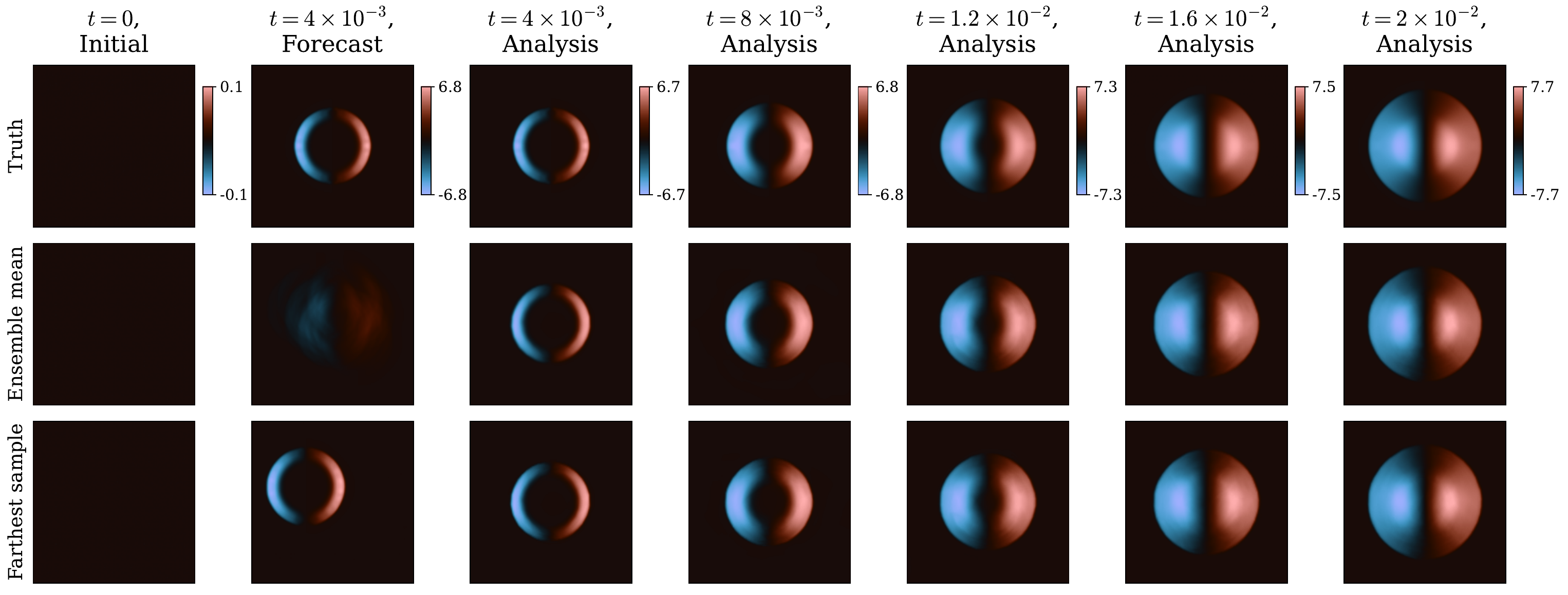}
\caption{
Velocity fields ($x$-component) for the 2D blast wave over the full assimilation window, demonstrating effective correction after the first neural EnKF update and sustained agreement with the reference solution thereafter. 
Rows show the reference solution, the ensemble mean (statistical state estimate), and the farthest ensemble member.
Columns correspond to the initial static condition ($t=0$), the forecast states prior to the first DA update ($t = 4.0\times10^{-3}$), and the analysis states after five successive DA steps up to the final time ($t = 2.0\times10^{-2}$).
}
\label{fig:bw_u_analysis}
\end{figure}

\begin{figure}[!htb]
\centering
\includegraphics[width=0.99\textwidth]{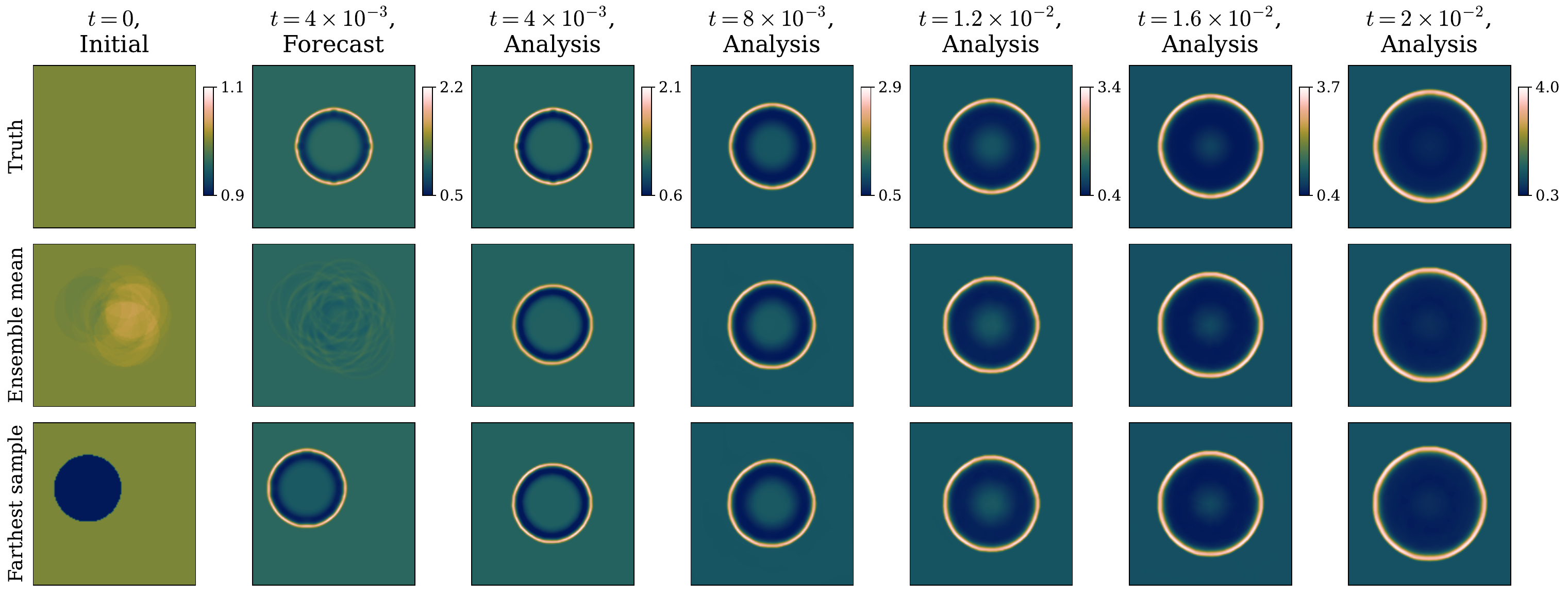}
\caption{Density fields for the 2D blast wave over the full assimilation window, demonstrating effective correction after the first two neural EnKF updates and sustained agreement with the reference solution thereafter. 
Rows show the reference solution, the ensemble mean (statistical state estimate), and the farthest ensemble member. 
Columns correspond to the initial condition ($t=0$), the forecast state prior to the first DA update ($t = 4.0\times10^{-3}$), and the analysis states after five successive DA steps up to the final time ($t = 2.0\times10^{-2}$).}
\label{fig:bw_rho_analysis}
\end{figure}

The neural EnKF also demonstrates robust DA performance in the 2D blast wave case, as illustrated in Figs.~\ref{fig:bw_p_analysis}--\ref{fig:bw_rho_analysis}, confirming its applicability to higher-dimensional shock-laden flows. Specifically, Fig.~\ref{fig:bw_p_analysis} shows the pressure fields over the full assimilation window, where the ensemble mean and the farthest ensemble member after each neural EnKF update are compared against the reference solution. 
Here, the farthest ensemble member is defined in the same manner as in the Sod shock tube problem.
Prior to the first DA update ($t = 4\times10^{-3}$), the forecast state exhibits noticeable uncertainty in the blast location and radius, leading to a smeared pressure discontinuity in the ensemble mean and a pronounced structural deviation in the farthest member. As a realization located at the edge of the ensemble distribution, this member provides a stringent test of the neural EnKF update. Following the first update, its pressure field undergoes substantial structural realignment, with the blast location and radius rapidly adjusted toward the reference solution, and the ensemble mean exhibits a similarly sharp correction.
As DA proceeds, subsequent updates further refine both the ensemble mean and the farthest member while preserving the discontinuity structure. Through the final time, the pressure fields remain closely aligned with the reference evolution, without introducing spurious oscillations.

Similarly, the velocity ($x$-component) and density fields, shown in Figs.~\ref{fig:bw_u_analysis} and~\ref{fig:bw_rho_analysis}, demonstrate effective structural correction after the first neural EnKF update. Although the ensemble mean remains slightly more diffused than the reference solution immediately after this initial update, the primary uncertainty in the discontinuity structure is substantially reduced. By the third DA step, both the ensemble mean and the farthest ensemble member become nearly indistinguishable from the reference solution. Subsequent updates provide only minor refinements while preserving the recovered discontinuity structure.

\begin{figure}[!htb]
\centering
\includegraphics[width=0.99\textwidth]{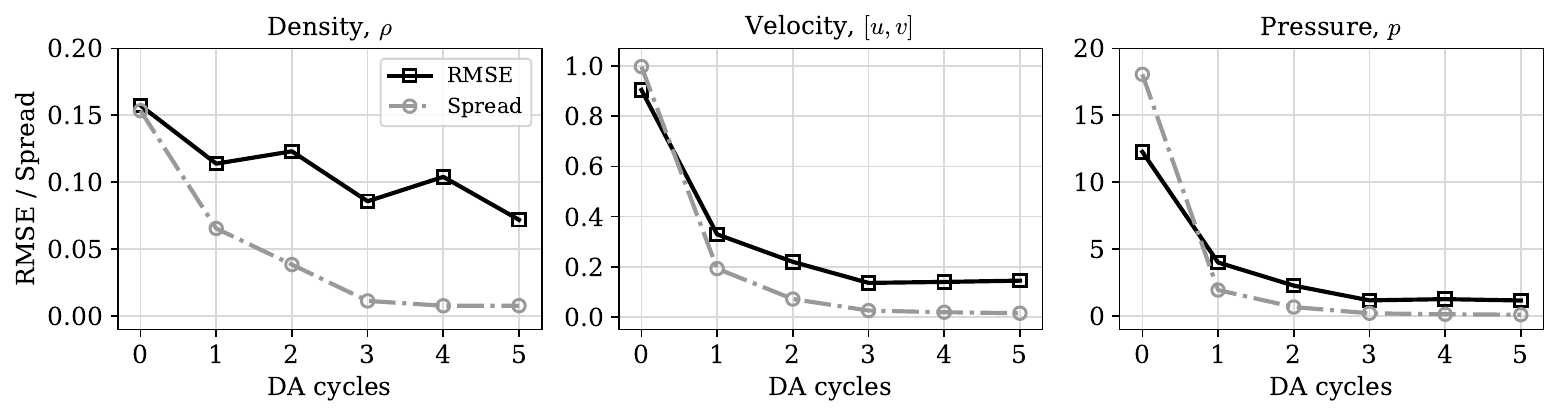}
\caption{Evolution of RMSE and ensemble spread over successive DA cycles for the 2D blast wave. 
Pressure and velocity exhibit a rapid reduction in both metrics after the first DA cycle, followed by sustained low levels. 
Density shows a comparable trend in ensemble spread, while its RMSE decreases more gradually and remains relatively low throughout the assimilation window. 
Cycle~0 corresponds to the forecast ensemble prior to the first DA update ($t = 4 \times 10^{-3}$), and cycles~1--5 denote analysis ensembles after successive DA steps.}
\label{fig:bw-rmse-spread}
\end{figure}

To further quantify the DA performance, 
Fig.~\ref{fig:bw-rmse-spread} presents the evolution of RMSE and ensemble spread over successive DA cycles. 
Consistent with the field-level comparisons above, the pressure field exhibits a sharp reduction in both RMSE and ensemble spread after the first DA cycle, followed by sustained low-error and low-uncertainty levels with a continued gradual decrease. 
The velocity field displays a comparable trend.  
The density field follows a similar overall pattern but with a more gradual reduction in the estimation error. 
Owing to the substantially smaller initial density contrast between the interior and exterior of the blast compared with the pressure jump, the density RMSE is already low prior to the first DA update and remains consistently low throughout the assimilation window. 
The relatively weak sensitivity of density to pressure observations may also contribute to this behavior.

\section{Conclusion}
\label{sec:conclusion}
This study considers DA for compressible flows with shocks and other types of discontinuities. We show that the EnKF fails in this problem due to a multimodal forecast distribution induced by uncertain shock locations. Such multimodality fundamentally violates the Gaussian assumptions underlying the EnKF update, leading to spurious oscillations and nonphysical artifacts.

To address this limitation, we introduce the neural EnKF framework, which reformulates the assimilation problem by performing updates in neural network parameter space rather than in physical state space. While representing individual physical states with neural networks is conceptually straightforward, the central difficulty lies in constructing an ensemble in neural space whose statistical structure is suitable for stable EnKF updates.
We resolve this issue through a nearest-neighbor chain training\slash transfer learning strategy. By ordering the training sequence according to physical-space similarity, the method promotes a consistent relationship between variations in neural-network parameters and variations in the underlying physical states. This alignment yields a structured neural-space ensemble that mitigates the adverse effects of multimodality observed in physical space and facilitates stable EnKF updates. Numerical experiments across multiple shock-dominated test cases demonstrate that the neural EnKF achieves robust, structure-preserving DA performance while substantially reducing spurious oscillations.

Several important directions warrant further investigation. 
First, although the chain-based training strategy is empirically effective, a rigorous theoretical characterization of the induced ensemble geometry in the neural space remains to be developed.
Second, extending the framework to large-scale 3D shock-laden flows will require further advances in computational and training efficiency, improved ensemble management, and scalable DA strategies such as localization~\cite{ott2004local,hunt2007efficient,tong2023localized}.
Third, the method relies on identifying a nonlinear transformation to a function representation that captures the smoothly varying features of the ensemble members and performing the analysis step in this representation; we have achieved this here with neural networks, but other approaches are possible.

\section*{Acknowledgments}
This work was supported by the AFOSR/AFRL Center of Excellence in Assimilation of Flow Features in Compressible Reacting Flows (Grant No. FA9550-25-1-0011).
Matthias Morzfeld is supported by the U.S. ONR Grant N000142512298.
The authors gratefully acknowledge Prof. Kevin Wang, Dr. Xuning Zhao, and Dr. Wentao Ma for their assistance with the \texttt{M2C} solver. 
The authors also thank Rahul Jain, Ruben Fernandez, and Prof. Venkat Raman for valuable discussions on the application context related to rotating detonation engines.
This work used computational resources provided by the Advanced Cyberinfrastructure Coordination Ecosystem: Services \& Support (ACCESS) program through allocation PHY250246, and was carried out on Anvil at Purdue University.

\section*{Code availability}
The source code is publicly available at \url{https://github.com/xu-hui-zhou/NeuralDA}.

\appendix
\section{Neural network architectures and training details}
\label{app:nn-arch}

\subsection{Neural network architectures}
Table~\ref{tab:nn-details} summarizes the neural network architectures used in the numerical experiments. In all cases, each ensemble member is represented by a fully connected network that maps spatial coordinates to the corresponding flow variables. ReLU activations are employed in the hidden layers, while the final layer uses no activation for velocity components and a softplus activation for density and pressure to enforce positivity.

The selected architectures were chosen to provide sufficient expressive capacity to accurately represent the flow states encountered throughout the DA process. In practice, this capability can be assessed prior to the DA experiments using representative snapshots sampled from the forecast trajectories of a small number of ensemble members.
In the 1D experiments (inviscid Burgers’ equation and the shock tube case), the networks are over-parameterized relative to the physical state dimension, a deliberate choice to ensure accurate representation of sharp structures while testing stability in high-dimensional weight space. In contrast, for the 2D blast-wave problem, the number of trainable parameters is already smaller than the corresponding physical state dimension. In practical 3D applications, where the discretized state dimension grows rapidly with spatial resolution, the parameter dimension need not scale proportionally, and we can therefore expect even greater gains in parameter efficiency.

\begin{table}[tb]
\caption{
Neural network architectures used in the numerical experiments.
}
\centering
\renewcommand{\arraystretch}{1.3}
\renewcommand\cellalign{cc}
\renewcommand\theadalign{cc}
\begin{tabular}{p{4.8cm} c c c}
\toprule[1.2pt]
 & 
Burgers' equation
& 
Sod shock tube
& 
Blast wave\\
\midrule

Input variables
& $x$
& $x$
& $x, y$ \\

Output variables
& $u$
& $\rho, u, p$
& $\rho, u, v, p$ \\
\midrule

\# input neurons
& 1
& 1
& 2 \\

\# hidden layers
& 4
& 4
& 6 \\

\# neurons per hidden layer
& 50
& 64
& 100 \\

\# output neurons
& 1
& 3
& 4 \\

Hidden-layer activation
& ReLU
& ReLU
& ReLU \\

Last-layer activation
& -
& Softplus ($\rho, p$ only)
& Softplus ($\rho, p$ only) \\
\midrule

\# trainable parameters
& 7{,}801
& 12{,}803
& 51{,}204 \\

Physical state dimension 
& 400 ($= 1 \times 400$)
& 1{,}200 ($= 3 \times 400$)
& 65{,}536 ($= 4 \times 128^2$) \\

\bottomrule[1.2pt]
\end{tabular}
\label{tab:nn-details}
\end{table}

We also experimented with two alternative architecture choices. 
First, we tested deeper and wider variants of the currently selected architectures and observed similar DA performance for the Sod shock tube and blast wave cases.
Second, we explored representing different flow variables using separate neural networks for each ensemble member. For example, in the Sod shock tube case, separate networks were used to represent density, velocity, and pressure individually. However, this led to less stable DA updates because the analysis could not reliably preserve shared structures across variables, such as maintaining consistent shock locations. Consequently, the final framework employs a single neural network to jointly represent all flow variables for each ensemble member.

\subsection{Training details}

For the forecast ensemble member $\mathbf{z}_i^{\mathrm f}$, the corresponding NN parameters $\bm{\theta}_i^{\mathrm f}$ are obtained by minimizing the reconstruction loss
\begin{equation}
\mathcal{L}(\bm{\theta}_i^{\mathrm f})
=
\frac{1}{n_x}
\sum_{j=1}^{n_x}
\left\|
\mathsf{F}_{\mathrm{NN}}(\bm{\theta}_i^{\mathrm f};\bm{x}_j)
-
\mathbf{z}_i^{\mathrm f}(\bm{x}_j)
\right\|_2^2,
\end{equation}
where $n_x$ denotes the number of spatial grid points in the computational domain and $\|\cdot\|_2$ denotes the Euclidean norm. The optimization is performed using the Adam optimizer~\cite{kingma2014adam}, and training is terminated once the reconstruction loss falls below a prescribed threshold, typically of order $10^{-6}$.

\begin{figure}[!htb]
\centering
\includegraphics[width=0.99\textwidth]{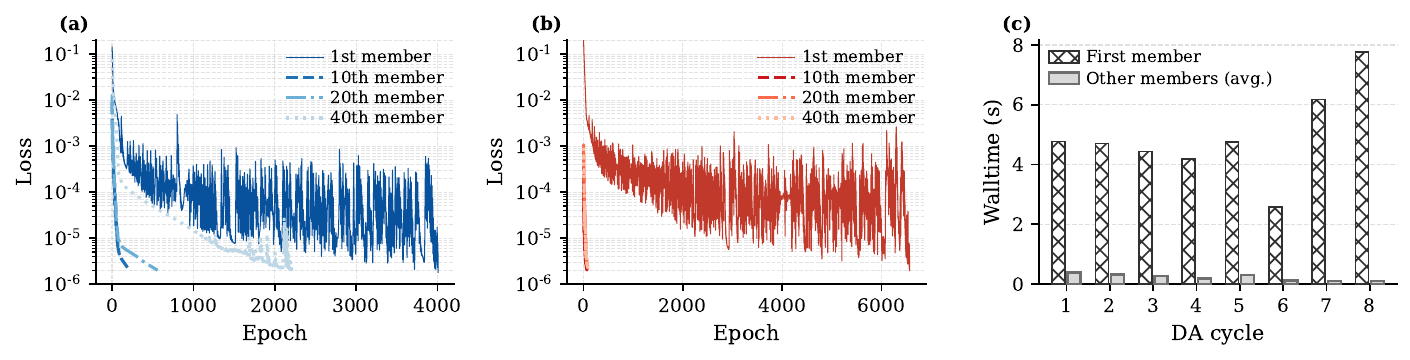}
\caption{Training convergence histories and wall-clock times for the Sod shock tube case: (a,b) training loss histories for selected ensemble members during the first and final DA cycles, respectively; and (c) training wall-clock times for each DA cycle, comparing the first ensemble member in the nearest-neighbor chain with the average over the remaining 39 ensemble members and demonstrating the significant reduction in training cost enabled by warm-start initialization in the chain-based training strategy. Here, the first ensemble member in the chain is trained using a learning rate of $10^{-2}$, while the remaining ensemble members use a learning rate of $10^{-3}$. All reported wall-clock times correspond to single-core CPU execution.}
\label{fig:training_summary}
\end{figure}

When applying this training procedure to the full forecast ensemble, we use the nearest-neighbor chain training strategy (see main text) to promote coherence between neural representations across ensemble members. This strategy also significantly accelerates NN training convergence through warm-start initialization, as illustrated in Fig.~\protect\ref{fig:training_summary}. Specifically, for the Sod shock tube problem, the convergence histories in panels (a) and (b) show that ensemble members later in the chain converge substantially faster than the first ensemble member trained from random initialization. This acceleration becomes even more pronounced in panel (b) during the final DA cycle, where later ensemble members converge within substantially fewer training epochs. This behavior is expected because the ensemble spread typically decreases as DA progresses, leading to greater similarity between neighboring ensemble members in the chain. Panel (c) further demonstrates the resulting reduction in NN training wall-clock times across DA cycles. Similar training behavior is also observed for the other test cases considered in this work.

\section{EnKF for the shock tube case}
\label{app:sod-enkf}

\begin{figure}[!htb]
\centering
\includegraphics[width=0.99\textwidth]{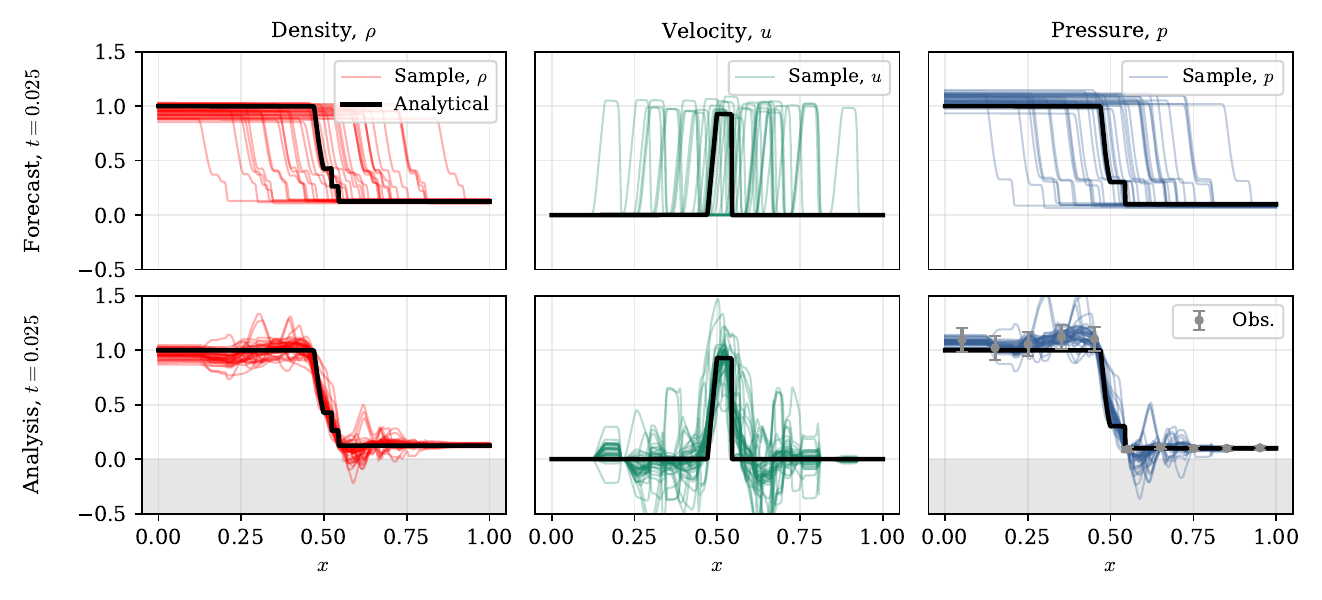}
\caption{EnKF performance for the shock tube problem at the first DA step ($t = 0.025$), showing strong oscillations and the emergence of nonphysical states. The shaded regions indicate the nonphysical range ($\rho < 0$ and $p < 0$).}
\label{fig:sod-posterior-enkf}
\end{figure}

Fig.~\ref{fig:sod-posterior-enkf} demonstrates that the analysis ensemble produced by the EnKF exhibits pronounced oscillations across the domain in the shock tube problem. These oscillations severely distort the solution structure, leading to nonphysical states with negative density and pressure and ultimately causing the forward solver to fail in the subsequent DA cycle.

\section{Pressure-density sensitivity in the shock tube case}
\label{app:sod-sensitivity}

To clarify why the density field recovers more slowly under pressure-only observations in the shock tube problem, we examine the forward propagation of uncertainty in the initial density. 
Here, uncertainty is introduced in the left-state density $\rho_L$,
while all other initial-condition parameters, including $p_L$, $\rho_R$, $p_R$, and the diaphragm location $x_d$, are fixed at their reference values (see Eq.~\eqref{eq:sod_ic}). 
This controlled setup isolates the effect of density perturbations and allows us to assess how uncertainty in $\rho_L$ propagates through the flow variables in the absence of DA.

\begin{figure}[!htb]
\centering
\includegraphics[width=0.99\textwidth]{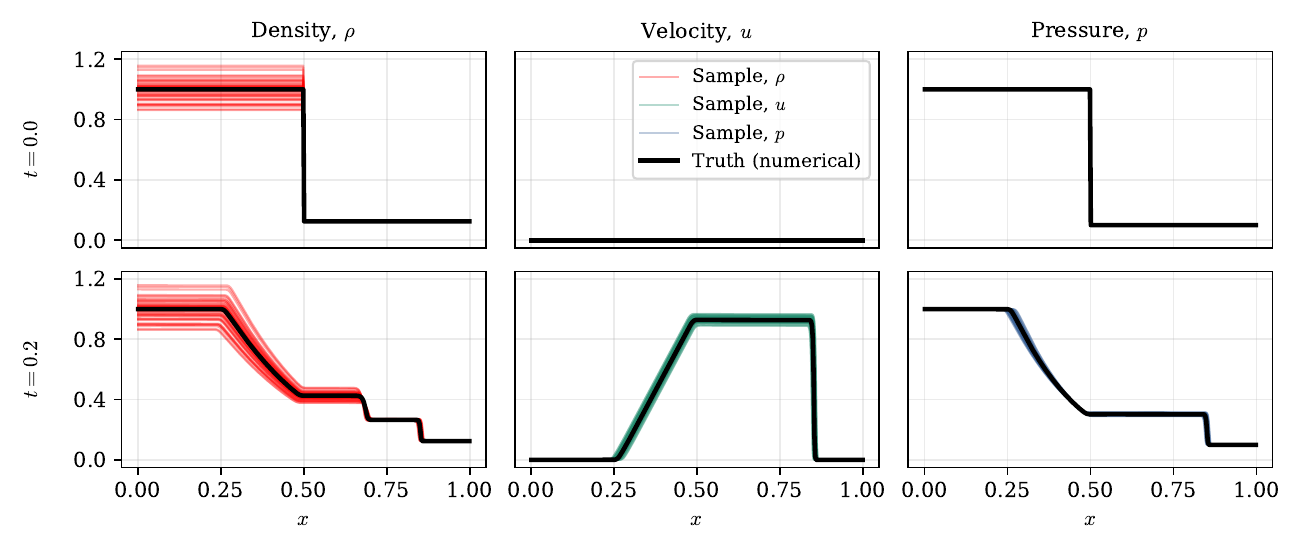}
\caption{Forward propagation of uncertainty in the left-state density $\rho_L$ for the shock tube problem, illustrating its weak influence on the pressure field. The ensemble is evolved from $t=0.0$ to $0.2$ without DA. Columns show density $\rho$, velocity $u$, and pressure $p$; rows correspond to the initial and final times.}
\label{fig:sod-sensitivity}
\end{figure}

As shown in Fig.~\ref{fig:sod-sensitivity}, the ensemble exhibits noticeable variability in the density field at $t=0$, confined to the left state where the perturbation is introduced. After evolution to $t=0.2$, the density profiles continue to display visible spread, whereas the corresponding pressure profiles remain tightly clustered across the ensemble.
To quantify this uncertainty propagation, the ensemble spread of the pressure field at $t=0.2$ is approximately $0.01$, compared with the imposed standard deviation of $0.1$ for the injected density perturbation. This suggests that perturbations in $\rho_L$ are only weakly transmitted to the pressure field over the time interval considered.

\section{Parametric study for the shock tube problem}
\label{app:sod-parametric}
To assess the robustness of the neural EnKF under different DA configurations, we conduct a parametric study for the shock tube problem. Specifically, we examine the sensitivity of the DA performance to variations in observation noise level, number of observation locations, and observation interval, while keeping all other numerical and DA settings fixed.

The configurations considered are summarized in Table~\ref{tab:SodMulExpSetup}. The baseline corresponds to the setup used in the main text. Experiments A1--A2 vary the observation noise level. Experiments B1--B2 alter the number of observation locations, equivalently changing the observation spacing $\Delta x_{\mathrm{obs}}$. Experiments C1--C2 vary the observation interval $\Delta t_{\mathrm{obs}}$, leading to different numbers of DA cycles over the entire simulation window ($T=0.2$).

\begin{table}[!htb]
\caption{
DA configurations used in the parametric study of the shock tube problem.
The baseline corresponds to the setup used in the main text.
Experiments A1--A2 vary the observation noise level;
B1--B2 vary the number of observation locations, thereby changing the observation spacing $\Delta x_{\mathrm{obs}}$;
and C1--C2 vary the observation interval $\Delta t_{\mathrm{obs}}$, leading to different numbers of DA cycles.
Parameters that differ from the baseline are highlighted in \textbf{bold}.}
\centering
\setlength{\tabcolsep}{12pt}
\renewcommand{\arraystretch}{1.4}
\begin{tabular}{l | c | c c | c c}
\hline
 & Obs.\ noise
 & \# Obs.\ loc.
 & $\Delta x_{\mathrm{obs}}$
 & \# DA cycles
 & $\Delta t_{\mathrm{obs}}$ \\
\hline
Baseline
& 5\%
& 10
& 0.1
& 8
& 0.025 \\
\hline
Exp.~A1
& \textbf{1\%}
& 10
& 0.1
& 8
& 0.025 \\
Exp.~A2
& \textbf{10\%}
& 10
& 0.1
& 8
& 0.025 \\
\hline
Exp.~B1
& 5\%
& \textbf{5}
& \textbf{0.2}
& 8
& 0.025 \\
Exp.~B2
& 5\%
& \textbf{20}
& \textbf{0.05}
& 8
& 0.025 \\
\hline
Exp.~C1
& 5\%
& 10
& 0.1
& \textbf{4}
& \textbf{0.05} \\
Exp.~C2
& 5\%
& 10
& 0.1
& \textbf{16}
& \textbf{0.0125} \\
\hline
\end{tabular}
\label{tab:SodMulExpSetup}
\end{table}

\begin{figure}[!htb]
\centering
\includegraphics[width=0.99\textwidth]{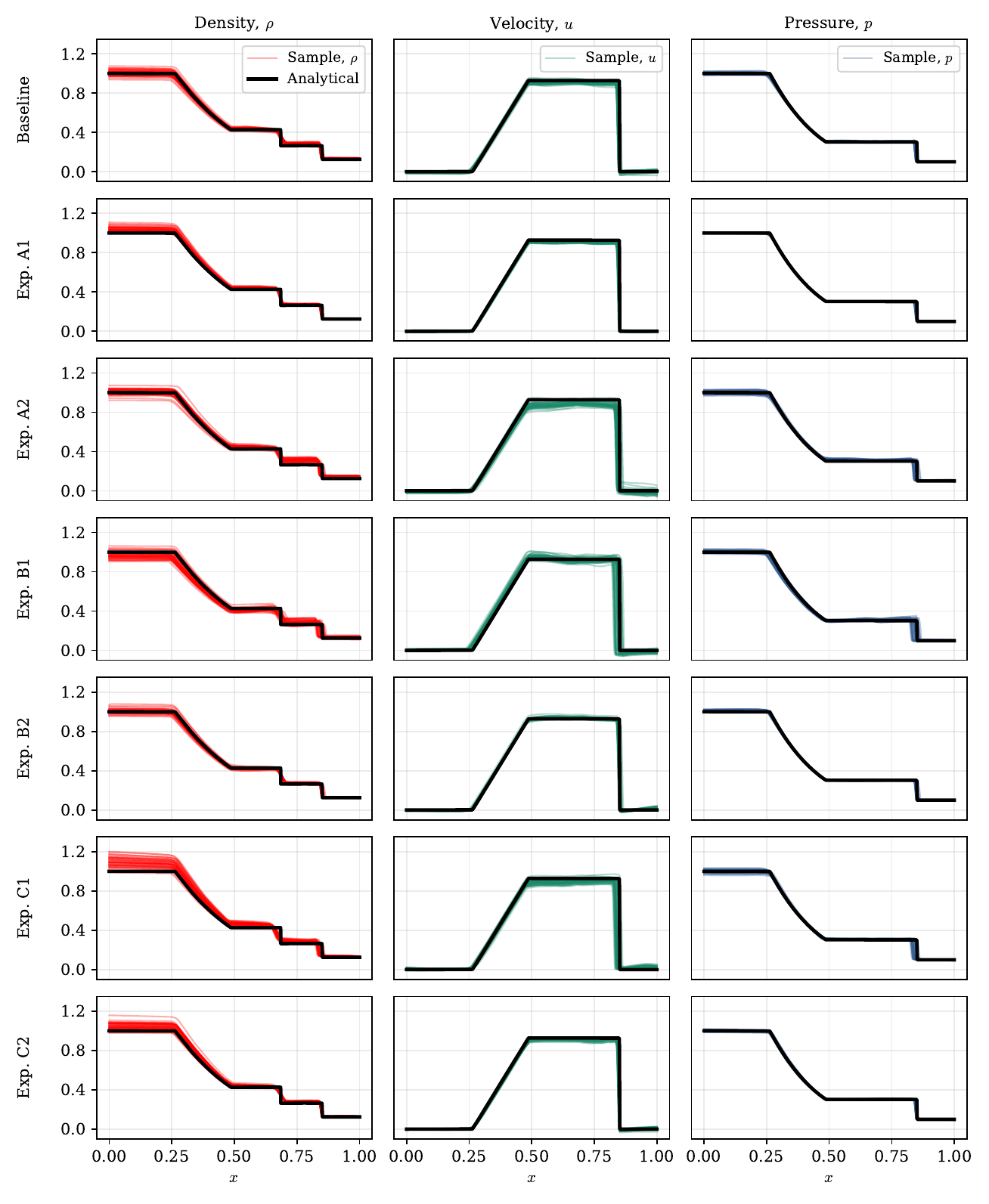}
\caption{
Final-time ($t=0.2$) analysis ensembles for the shock tube problem under the DA configurations summarized in Table~\protect\ref{tab:SodMulExpSetup}. Columns show density $\rho$, velocity $u$, and pressure $p$. Rows correspond to the baseline case used in the main text; Experiments A1--A2 (low and high observation noise); Experiments B1--B2 (reduced and increased numbers of observation locations); and Experiments C1--C2 (lower and higher observation frequency).}
\label{fig:sod-parametric}
\end{figure}

The final-time analysis ensembles obtained under different DA configurations are shown in Fig.~\ref{fig:sod-parametric}. Overall, the neural EnKF exhibits robust performance across all tested settings, consistently recovering the dominant flow structures. The sharpness of discontinuities and the ensemble spread, however, vary systematically with the strength of the observational constraints.

Specifically, stronger observational constraints---reduced observation noise (Exp.~A1), a larger number of observation locations (Exp.~B2), or more frequent assimilation (Exp.~C2)---lead to smaller ensemble spread and sharper reconstruction of discontinuities, particularly the shock. In these cases, the analysis ensembles are more tightly concentrated around the analytical solution, indicating improved estimation accuracy and reduced uncertainty.
Conversely, weaker constraints---higher noise levels (Exp.~A2), fewer observation locations (Exp.~B1), or less frequent assimilation (Exp.~C1)---result in larger ensemble spread and less sharply localized transitions near discontinuities. Among these factors, using fewer observation locations (Exp.~B1) produces the most noticeable degradation, highlighting the difficulty of accurately localizing shocks when spatial information is insufficient.

Across all configurations, pressure remains the most robustly recovered variable due to direct observation. Velocity follows closely owing to its strong dynamical coupling with pressure, whereas density exhibits greater sensitivity to the DA configuration, reflecting its weaker constraint by pressure observations alone, as discussed in~\ref{app:sod-sensitivity}.

\end{document}